\documentclass[aps,prb,twocolumn,showpacs,floatfix,eqsecnum,superscriptaddress]{revtex4}
\usepackage{graphicx}
\usepackage{bm}
\usepackage{dcolumn}
\def\he4{$^4$He}
\def\Am3{\AA$^{-3}$}
\def\beq{\begin{equation}}
\def\eeq{\end{equation}}

\begin{document}
\def\he4{$^4$He}
\def\Am2{\AA$^{-2}$}
\def\Am3{\AA$^{-3}$}
\def\beq{\begin{equation}}
\def\eeq{\end{equation}}
\title{Worm Algorithm and Diagrammatic Monte Carlo: A New Approach\\
to Continuous-Space Path Integral Monte Carlo
Simulations}

\author{M. Boninsegni}
\affiliation{Department of Physics, University of Alberta,
Edmonton, Alberta T6G 2J1}

\author{N.V. Prokof'ev}
\affiliation{Department of Physics, University of Massachusetts,
Amherst, MA 01003, USA}
\affiliation{BEC-INFM, Dipartimento di Fisica,
Universita di Trento, Via Sommarive 14, I-38050 Povo, Italy}
\affiliation{Russian Research Center ``Kurchatov Institute'',
123182 Moscow, Russia}

\author{B.V. Svistunov}
\affiliation{Department of Physics, University of Massachusetts,
Amherst, MA 01003, USA}
\affiliation{Russian Research Center ``Kurchatov Institute'',
123182 Moscow, Russia}


\begin{abstract}
A detailed description is provided of  a new Worm Algorithm,
enabling the accurate computation  of thermodynamic properties  of
quantum many-body systems in continuous space, at finite
temperature. The algorithm is formulated  within the general Path
Integral Monte Carlo (PIMC) scheme, but also allows one to perform
quantum simulations in the grand canonical ensemble, as well as to
compute off-diagonal imaginary-time correlation functions, such as
the Matsubara Green function, simultaneously with
diagonal observables. Another important innovation consists of the
expansion of the attractive part of the pairwise potential energy
into elementary (diagrammatic) contributions, which are then
statistically sampled. This affords a complete microscopic account
of the long-range part of the potential energy, while keeping the
computational complexity of all updates independent of the size of
the simulated system. The computational scheme allows for
efficient calculations of the superfluid fraction and off-diagonal
correlations in space-time, for system sizes which are orders of
magnitude larger than those accessible to conventional PIMC. We
present illustrative results for the superfluid transition in bulk
liquid \he4 in two and three dimensions, as well as the
calculation of the chemical potential of  {\it hcp} \he4.
\end{abstract}

\maketitle

\section{Introduction}
It is now twenty years since Ceperley and Pollock (CP) carried out
the first Path Integral Monte Carlo (PIMC) simulation of the
superfluid transition of liquid \he4.\cite{ceperley86} Albeit
restricted to a system of 64 \he4 atoms with periodic boundary
conditions, that study demonstrated the feasibility of {\it ab
initio} numerical studies of quantum many-body systems, the mass
of the particles and the interaction potential being the sole
input to the calculation.

The PIMC method, in the form developed by CP (henceforth referred
to as ``conventional"), has since played a major role in the
theoretical  investigation of quantum many-body systems. Not only
has it provided quantitative results for a wide range of physical
systems, it has also shaped, to some extent, our qualitative
understanding  of such phenomena as superfluidity (SF) and Bose
condensation, at the microscopic level. At least for Bose systems,
PIMC is the {\it only} presently known method capable of
furnishing in principle {\it exact} numerical estimates of
physical observables at finite temperature ($T$), including the
superfluid  ($\rho_s$) and condensate ($n_\circ$) fractions.
Moreover, despite the notorious {\it sign} problem, that has so
far made it impossible to obtain equally high quality results for
Fermi systems, PIMC proves a valid option in this case as well,
allowing one to obtain approximate estimates of accuracy at least
comparable to that afforded by the other leading
methods.\cite{ceperley95,ceperley96}

It thus seems reasonable to regard PIMC as a realistic
option to investigate ever more complex quantum many-body systems,
and it makes sense to try and overcome its most important present
limitations.  Aside from the above-mentioned {\it sign} problem,
which we do {\it not} discuss in this paper, the main bottleneck
of the current PIMC technology is inarguably the maximum system
size (i.e., number $N$ of particles) for which accurate estimates
can be obtained, in a reasonable amount of computer time. Specifically,
the computational effort required to study properties that most directly depend 
on particle indistinguishability, is observed to scale prohibitively with $N$.

For example, the superfluid fraction $\rho_s$ is obtained in a
PIMC simulation of bulk condensed matter, by
means of the so-called winding number estimator,\cite{pollock87}
which can only take on a nonzero value if long permutation cycles
of identical particles occur. In conventional PIMC, the frequency
with which such cycles are sampled, is an {\it exponentially}
decreasing function of $N$. 
For this reason, and in spite
of (at least) a hundredfold increase in computer speed,\cite{pt} since the
pioneering work of Ref. \onlinecite{ceperley86} it has not proven
possible to obtain estimates of the superfluid fraction  in bulk
liquid \he4 for finite systems of more than $N$=64 particles. 
Besides the unfavorable scaling 
of computing resources as a function of $N$, another major issue that
this entails
is the difficulty of assessing reliably whether  the observed absence of long 
permutation cycles reflects a
genuine physical effect, or merely lack of ergodicity of the path
sampling scheme.\cite{bernu04}

How important is the above size limitation ? For most observables
diagonal in the coordinate representation, one can often approach
surprisingly closely the thermodynamic  limit by
simulating as few as $\sim$ 30 particles, especially for systems
characterized by short-ranged,  Lennard-Jones-type interactions.
On the other hand, an  accurate {\it quantitative}
characterization of  the superfluid transition (including the
calculation of the transition temperature $T_{\rm c}$) can only be
obtained via  finite-size scaling analysis of results for
$\rho_s(T)$ and/or $n_\circ(T)$. The reliability of this procedure
crucially hinges on the availability of  data for large systems of
significantly different sizes. Attempts to estimate $T_{\rm c}$
for superfluid \he4, based on PIMC data for $\rho_s(T)$ for
systems of size $N$=64 and smaller, failed to yield quantitative
results.\cite{pollockrunge}

 But there are other  reasons, arguably
more important than the mere pursuit of numerical accuracy,
pointing to the importance and timeliness of extending by one or
two orders of magnitude the size of the systems  accessible to
PIMC. Quite generally, in order for a  scientific question to be
meaningfully addressed by numerical simulations, the size of the
simulated system should be greater than all
characteristic length scales affecting  the physics  of interest. This is
particularly important in the study of inhomogeneous phases of
matter, or fluids in confinement, or restricted geometries. An
example is provided by the study of helium fluid in porous glass,
such as Vycor; the diameter of a characteristic pore is of the
order of a few tens of \AA. Thus, a realistic calculation, at the typical
liquid helium density, requires that one be able to simulate a
system comprising several thousands of atoms. Other examples are
the numerical investigations of multicomponent systems, as well as
of grain boundaries, dislocations and other defects in quantum
solids, or of incommensurate phases of films of helium or {\it
para}-hydrogen adsorbed over substrates such as graphite.

Over the past two decades, there has been relatively little
experimentation with approaches to PIMC simulations differing in
some  important aspects from the conventional one of CP,
thoroughly described in Ref. \onlinecite{ceperley95}. As mentioned
above, in conventional PIMC the simulation of properties that are
most directly affected by quantum statistics (i.e., by
many-particle permutations), suffers from a very unfavorable
scaling of required computer time with system size.  This hurdle
seems difficult to conquer within conventional PIMC, and more
generally within any Monte Carlo scheme formulated in the
canonical ensemble, in which  the winding number becomes
``topologically locked'' in the $N\to \infty$
limit.\cite{ceperley95}

On the other hand, the same hurdle has been completely overcome in
Quantum Monte Carlo (QMC) simulations of lattice models. A lattice
Path Integral scheme based on an alternative sampling approach,
known as {\it worm algorithm} (WA), \cite{worm} has been
demonstrated to allow for efficient calculations of winding
numbers and of the one-particle Green function $G$, for systems of
as many as $\sim$ 10$^6$ particles.\cite{prokofev04} It is
particularly useful for the studies of critical phenomena since it
does not suffer from the critical slowing down problem
\cite{worm2001} present in other local-update schemes.

The WA has been recently extended to the study of systems in
continuous space;\cite{worm1} it has first been shown to afford
the simulation of the superfluid transition in liquid \he4 in two
dimensions, for systems comprising as many as 2500 particles,
i.e., about 100 times greater than those accessible to
conventional PIMC. Subsequently, it has been applied to the study
of Bose condensation in crystalline \he4,\cite{superglass}  as
well as to the investigation of superfluid properties of {\it
para}-Hydrogen droplets.\cite{mezzacapo}

In all of these applications, the WA has provided accurate
numerical results, simply not obtainable with any other existing
method. It need be stressed, however, that the WA is not merely
about doing large system sizes (important as this is); it is also
the first  {\it grand canonical} QMC method with local updates to
incorporate in full quantum statistics. It affords the exact
computation of imaginary-time off-diagonal correlations, such as
the one-particle Matsubara Green function, that are not accessible
to conventional PIMC (nor to any other QMC technique in continuous
space).

In this paper, we provide a detailed description of this new,
powerful computational tool, which promises  to open novel avenues
to the theoretical exploration of strongly correlated many-body
system. The manuscript is organized as follows: in the next
section (\ref{simplest}), we describe the simplest implementation
of the WA, also introducing our nomenclature, configurational
space and data structure. In  section \ref{mgf}, we discuss in
detail the computation of the Matsubara Green function, as well as
of its equal-time limit, namely the one-particle density matrix.
In section \ref{dia}, we describe an important enhancement of the
simple implementation, namely the expansion of (the attractive
part of) the potential energy of interaction among particles in
elementary (diagrammatic) contributions, that are then sampled by
a Monte Carlo method. This scheme, which falls in the general
category of  Diagrammatic Monte Carlo techniques,\cite{DMC} allows
for the full inclusion of the contribution of the potential
energy (for an important class of potentials) at a computational cost that is {\it independent} of the
size of the system. 

In Sec. \ref {add}, we provide some quantitative
information related to the specific utilization of the WA in
simulation studies of superfluid \he4. In Sec. \ref{resu},  we
offer a quantitative demonstration of the power of the WA, by
illustrating in detail our results for simulations of the
superfluid transition of liquid \he4 in two and three dimensions.
We outline our conclusions, and discuss outlook for future
applications of the WA, in Sec. \ref{conclusions}.

\section{Simplest Version}\label{simplest}

We begin with some basic notation. We assume for definiteness a
system of identical particles (in $d$ dimensions) obeying Bose
statistics.\cite{notebf} Let $m$ be the mass of each particle. The
system is enclosed in a cubic vessel of volume $V=L^d$, with
periodic boundary conditions in all directions (other geometries,
boundary conditions, and/or external forces require obvious and
minimal modifications which are standard for any QMC scheme).

We make from the outset the assumption of working in the {\it
grand canonical} ensemble; that is, the system is held in thermal
equilibrium with a heat reservoir at temperature $T=1/\beta$ (we
set $k_B$=1), with which it can exchange particles as well.
Consequently, in order to specify the thermodynamic state of the
system,  we need to assign the chemical potential $\mu$, which is
an input parameter in our computational scheme, just like the
temperature $T$. The number of particles $N$ is allowed to
fluctuate.

Let $\hat H$ be the (many-body) system Hamiltonian, which we
assume of the following form: \beq \hat H = -\lambda\sum_{i=1}^N\
\nabla_i^2 + \sum_{i < j} v(|{\bf r}_i-{\bf r}_j|) \; ,\eeq
where $\lambda=\hbar^2/2m$ and where $v$ is a pairwise interaction
potential that depends only of the relative distance between any
two particles.\cite{notev}
 Below, $R$  will always be used as a collective ``coordinate", representing positions
of all particles in the system, i.e., $R\equiv ({\bf r}_1,{\bf r}_2,\dots ,{\bf r}_N)$.

\subsection{Configurational Space}\label{cspace}
A fundamental aspect of the WA, which crucially distinguishes it
from conventional PIMC and from all existing QMC methods in the
continuum, is that it operates in an extended configurational
space, containing both closed world line configurations
(henceforth referred to as $Z$- or diagonal configurations), as
well as configurations containing one open world line (worm). The
$Z$-configurations contribute to the partition function, whereas
those with an open world line contribute to the one-particle
Matsubara Green function; in the following, the latter  will be
referred to as $G$- (or, off-diagonal) configurations. As we shall
see, all topologically non-trivial modifications of world lines
occur in the off-diagonal configurational space (or, $G$-sector).
The sampling process allows for transitions from the $G$- to the
$Z$-sector (by closing, or removing the existing open world line)
and vice versa (by creating a new open world line, or by opening
an existing closed one). Expectation values of all physical
quantities of interest (with the exception of the Green function),
including particle and winding numbers, are only updated when the
random walk generates a diagonal configuration.

Next, we proceed to describe in detail the two sectors in which our configuration space is conceptually divided.

\subsubsection{The $Z$-sector}

{\it The $Z$-sector} of our configuration space, is nothing but the full configuration space
of conventional PIMC. It naturally emerges from the
Path Integral representation of the grand partition function $Z={\rm Tr}\,
e^{-\beta\hat K}$, where $\hat K = \hat H -\mu \hat N$.

Each $Z-${\it configuration} is a discrete
imaginary-time many-particle path, $X\equiv (R_0, R_1, R_2, \ldots ,
R_P)$, with $R_P\equiv R_0$ (except for a possible permutation of particle labels), representing the integrand in the
asymptotically exact (in the $P\to \infty$ limit) integral
decomposition of $Z$
\begin{equation}\label{zint}
Z  \approx \sum_{N=0}^\infty\ e^{\beta \mu N} \int  dX\
A(X,\varepsilon)\, {\rm e}^{-U(X)} \; , \label{start}
\end{equation}
where $dX\equiv dR_0 dR_1...dR_{P-1}$, $\varepsilon =\beta/P$, and where
\[
A(X,\epsilon) \equiv \prod_{j=0}^{P-1} \rho_F
(R_j,R_{j+1},\varepsilon)\; . \] In turn, $\rho_F$ is a product of
free-particle imaginary-time propagators, i.e., with an obvious
notation, \beq
\rho_F(R_j,R_{j+1},\epsilon)=\prod_{i=1}^N\rho_\circ({\bf
r}_{ij},{\bf r}_{i,j+1},\varepsilon) \eeq with \beq
\rho_\circ({\bf r},{\bf r^\prime},\epsilon) =
(4\pi\lambda\epsilon)^{-d/2}\ {\rm exp}\biggl [ -\frac{({\bf
r}-{\bf r^\prime})^2}{4\lambda\epsilon}\biggr ] \; . \eeq The
function $U$ in Eq. (\ref{zint}) incorporates correlations, both
in space and in imaginary time, arising from interactions among
particles. $U$ is chosen so that, in the $\varepsilon\to 0$ limit,
the distribution of discrete paths $X$ will asymptotically
approach the correct continuous limit. Several choices are
possible \cite{ceperley95} for $U$, but the simplest version of
the algorithm described in this section does not depend on its
particular form. In what follows, we refer to the product
$W(X)\equiv A(X,\varepsilon)\ {\rm e}^{-U(X)}$ as a {\it
configurational weight}.

Eq.~(\ref{start}) implies the following configuration space
structure: one has $N$ single-particle paths (world lines), labeled
$i=1,2,\ldots ,N$, propagating in the discretized imaginary  time interval
$[0,\beta]$ (specifically, $t_0=0$, $t_P=\beta$). Each world line consists of $P$
successively linked ``beads" (particle positions), labeled by the index of the
corresponding imaginary time ``slice", $j=0,\ldots ,P-1$. The $j$-th bead of
the $i$-th world line is positioned at ${\bf r}_{ij}$.

As a result  of $\beta$-periodicity, coupled with the physical
indistinguishability of particles, the $(P-1)$-st bead of each
world line must be linked to the zero-th bead of either the same,
or another world line. For both theoretical and practical (data
structure) purposes, it is advantageous to guarantee
$\beta$-periodicity automatically; to this aim, we regard world
line configurations as closed loops on a $(d+1)$-dimensional
surface of a $(d+2)$-dimensional $\beta$-cylinder, on which lie
$P$ equidistant (and equivalent) imaginary time ``hyperplanes"
(corresponding to the different time slices), labelled
$j=0,\ldots, P-1$. It should be noted that the total number of
world line loops defined on the $\beta$-cylinder can be different
from the total number of particles; this is because a single world
line which ``winds around" the imaginary time interval $l$ times
before returning to its initial position represents not just one
particle, but rather $l$ particles involved in the same exchange
cycle. The presence of such exchange cycles is essential, in order
to incorporate in the computational scheme the symmetry of the 
system with respect to particle permutations.

\subsubsection{The $G$-sector}
{\it The $G$-sector} of our configurational space comes from the
representation---analogous to that of the partition function,
Eq.~(\ref{start})---of the one-particle Matsubara Green function
\begin{equation}
G({\bf r}_1, {\bf r}_2, \tau)\; =\;  \langle \, {\cal T} \{ \hat
\psi({\bf r}_1,\tau)\ \hat \psi^\dagger({\bf r}_2,0) \}\, \rangle
\; \equiv \; { g({\bf r}_1, {\bf r}_2, \tau) \over Z} \; ,
\label{g}
\end{equation}
where $\langle \ldots \rangle$ denotes thermal averaging, ${\cal
T}$ is the time-ordering operator and $\hat \psi^\dagger ({\bf r},
\tau)$ and $\hat \psi({\bf r},\tau)$ are (Bose) particle creation
and annihilation operators in Matsubara representation.  The
structure of the integral representation of $g({\bf r}_1, {\bf
r}_2, \tau)$ is very similar to that of $Z$. In fact, the only
qualitative difference of a $G$-sector (``off-diagonal")
configuration from a diagonal one, is that the former contains a
{\it worm}, that is, a world line on a $\beta$-cylinder with two
ends---the ``head" and the ``tail"---corresponding to the Green
function annihilation and creation operators, respectively. The
two special beads at the open world line ends are named (for
historical reasons) {\it Ira} (${\cal I}$) and {\it Masha} (${\cal
M}$). Configurations in which ${\cal I}$ and ${\cal M}$ are
located in space-time at points $({\bf r}_{\cal I}, \tau_{\cal
I})$ and $({\bf r}_{\cal M}, \tau_{\cal M})$ contribute to $g({\bf
r}_{\cal I}, {\bf r}_{\cal M}, \tau_{\cal I}-\tau_{\cal M})$ with
the weight defined in accordance with Eq.~(\ref{start})
generalized to include the off-diagonal configuration sector.

Formally, the ensemble of WA configurations  corresponds to the
generalized partition function
\begin{equation}
Z_W=Z+Z'\; ,
\end{equation}
with
\begin{eqnarray}
Z'  =\,  C \! \sum_{j_{\cal I}, j_{\cal M} }  \int d{\bf r}_{\cal
I}\; d{\bf r}_{\cal M}\;  g({\bf r}_{\cal I}, {\bf r}_{\cal M}, \,
\varepsilon({j_{\cal I}} - {j_{\cal M}}))\; . ~~~ \label{ZW}
\end{eqnarray}
The value of dimensionless parameter $C$ only affects the
efficiency of the simulation, as $C$ controls the relative
statistics of $Z$- and $G$-sectors; for the moment, we leave it
undetermined, to come back to  it later on, when discussing
updates.

An important new feature that arises  when going from $Z$ to
$Z_W$, is that the number of continuous variables in $G$-sector
configurations is not constant, but rather varies from
configuration to configuration. This immediately points to
Diagrammatic Monte Carlo,\cite{DMC} as a general way to perform
updates whenever the number of variables to sample,  is itself
variable.

The sampling of paths $\{X_l \}$, is implemented within the WA {\it
exclusively} through a set of simple, {\it local} space-time updates involving
${\cal I}$  (or, ${\cal M}$). The particle number
becomes configuration- and time-dependent (there is one less
particle between ${\cal I}$ and ${\cal M}$, than in the rest of
the path). This clearly shows  how, by its very construction, the WA
opens up the possibility of working in
the {\it grand canonical} ensemble, with the chemical potential
$\mu$ being an input parameter.\cite{note1} Obviously, WA updates
can be combined with the conventional PIMC updates to have
the most flexible scheme.

In accordance with Eq.~(\ref{ZW}), the simplest estimators in WA
are [assuming that configurations are sampled from the probability
density $W(X)$]
\begin{equation}
\delta^{(Z)}\, =\; \left\{
\begin{array}{l}
1 \; , ~~ \mbox{if in}~Z\mbox{-sector}  \; , \\
0\; , ~~\mbox{if in}~G\mbox{-sector}\; ,\end{array} \right.
\label{delta_Z}
\end{equation}
\begin{equation}
\delta^{(G)}\, =\; \left\{
\begin{array}{l}
0 \; , ~~ \mbox{if in}~Z\mbox{-sector}  \; , \\
1\; , ~~\mbox{if in}~G\mbox{-sector}\; .\end{array} \right.
\label{delta_G}
\end{equation}
In the statistical limit, their Monte Carlo averages are
\begin{equation}
\langle \, \delta^{(Z)} \rangle_{\rm MC} \;  = \;  Z/Z_W \; ,
\end{equation}
\begin{equation}
\langle \, \delta^{(G)} \rangle_{\rm MC} \; =\; {C P\over Z_W}\,
\sum_{j=0}^{P-1}  \int\!  d {\bf r}_1\, d {\bf r}_2\;  g({\bf
r}_1, {\bf r}_2, \varepsilon j) \; ,
\end{equation}
In particular
\begin{equation}
{ \langle \, \delta^{(G)} \rangle_{\rm MC} \over \langle \,
\delta^{(Z)} \rangle_{\rm MC} } \; =\; C P\, \sum_{j=0}^{P-1}
\int\!   d {\bf r}_1\, d {\bf r}_2\;  g({\bf r}_1, {\bf r}_2,
\varepsilon j) \; . \label{2-11}
\end{equation}
The simplest estimator for $g({\bf r}, \tau)$, $\tau = \varepsilon
j$, is given by
\begin{eqnarray}\nonumber
 \langle \, \delta^{(G)}
\, \delta_{j,\, (j_{\cal I}-j_{\cal M})}\, \delta({\bf r}_1 - {\bf
r}_{\cal I})\, \delta({\bf r}_2-{\bf r}_{\cal M})\, \rangle_{\rm
MC} \; =\\ \, = \; {CP\over Z_W}\;  g({\bf r}_1, {\bf r}_2,
\varepsilon j) \; .~~~~~~~~~~~ \label{grj}
\end{eqnarray}
However, below we introduce a more elaborate scheme which allows
one to circumvent the problem of working with generalized
functions (typically solved by collecting statistics to
finite-size spatial bins at the expense of an additional
systematic error).

\subsection{Data Structure and Updates}
In this section,  we describe a set of ergodic local updates which sample
the extended configuration space, switching between the $Z$- and
$G$-sectors. Updates which change the number of continuous
variables in $X$, are arranged in complementary pairs, designed so as to satisfy
the requirement of detailed balance. General principles of balancing complementary
pairs can be found in Ref.~\onlinecite{DMC}. We have three pairs of updates altogether:
{\it Open/Close}, {\it Insert/Remove}, and {\it Advance/Recede}.
Only the {\it Swap} update in the list below does not fall in this
category, because it preserves the number of variables, i.e., it
is self-complementary. Proposed updates are either accepted or rejected based on the
Metropolis algorithm, according to the standard procedure.\cite{metropolis}

In order to keep the presentation simple, the sampling scheme described below is one in which
every update can be proposed, regardless of its applicability to the current configuration (for example,
the proposal to {\it Remove} the worm is allowed even if there is no worm in the current
configuration, in which case the proposed update will necessarily be rejected).
It should be understood, however, that standard sampling tricks can be used, whereby
only applicable updates are proposed, thereby enhancing the performance.\cite{note3}

\begin{figure}[t]
\vspace*{-0cm}
\includegraphics[bb=50 280 530 590, width=\columnwidth]{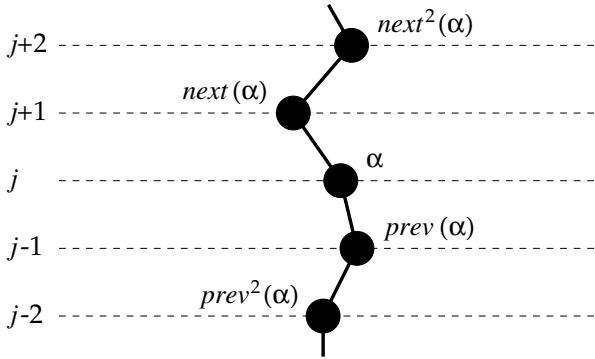}
 \caption{World line
beads on the $\beta$-cylinder and nearest neighbor associations
between them. } \label{fig1}
\end{figure}

To be specific, let us adopt the following data structure:
all beads are labeled, and each bead is linked to its
two world-line neighbors, the {\it next} and the {\it previous}
beads on the $\beta$-cylinder, see Fig. \ref{fig1}. It proves convenient to
introduce two functions, '$next$' and '$prev$', mapping each bead
onto its next and previous neighbor, respectively. Correspondingly,
$\sigma=next^m (\alpha)$ means that the bead $\sigma$ is the result
of the $m$-fold application of '$next$' to the bead $\alpha$;
likewise, $\alpha=prev^m (\sigma)$.

In order to have an efficient spatial addressing of beads, and thus a scalable
algorithm (i.e.,  one in which the number of operations required to perform updates
does not depend on either system size, nor temperature), we use nearest-neighbor
tables at all imaginary time slices.\cite{allen}
That is, for each time slice, the volume of the system is hashed into equal
microscopic ``bins", labeled by the discrete time label,
$j=0,1,2,\ldots , (P-1)$, and discrete radius vector, $\vec{\cal
R}$. [In practice, we set the bin volume, $\Omega$, to be of the
order of the volume per particle.] For each bin there is a list of
all beads contained in it.
Accordingly, addressing/searching
relevant beads is performed through nearest-neighbor tables and next/previous
links.

\begin{figure}[t]
\vspace*{-0cm}
\includegraphics[bb=50 250 515 670, width=\columnwidth]{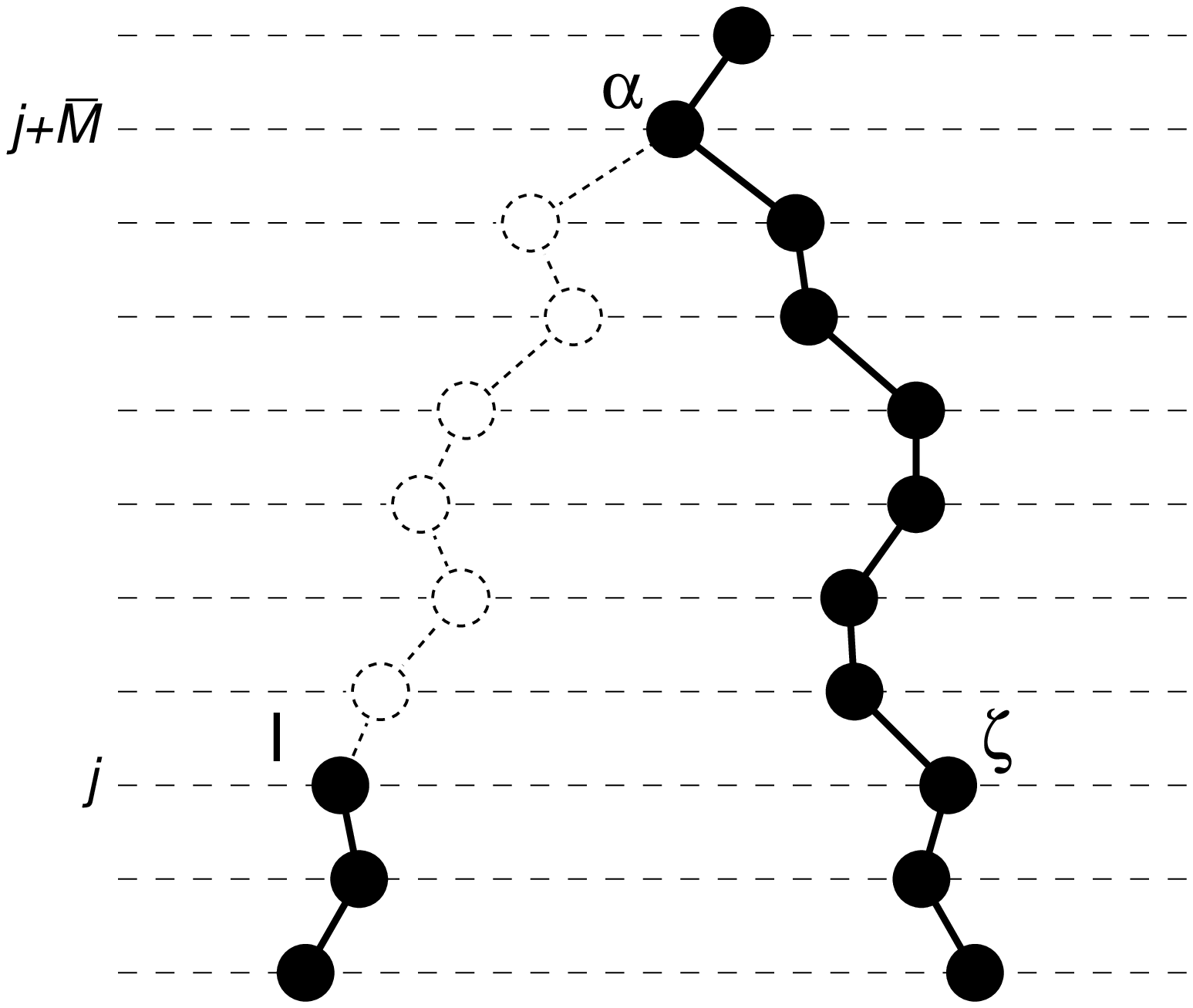}
\includegraphics[bb=50 250 515 680, width=\columnwidth]{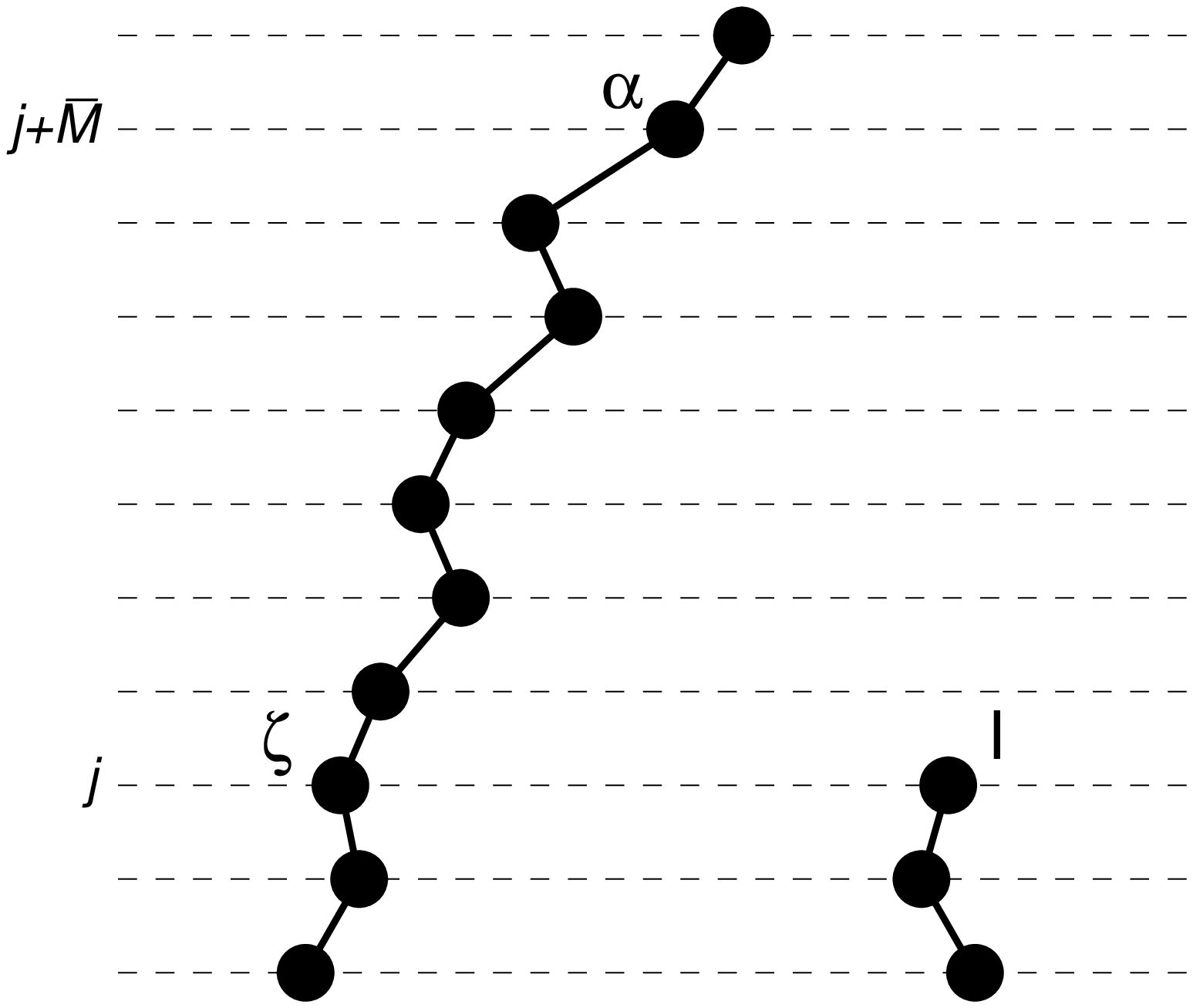}
 \caption{An illustration of the {\it Swap} update and the notation
 introduced in the main text. } \label{fig2}
\end{figure}

\indent  {\it (1a) Open}. The update is only possible if the
configuration is diagonal. A bead $\alpha$ is selected at
random. An integer number $M$ is selected at random within the
interval $[1,\bar{M}]$ with $\bar{M} < P$ being an arbitrary
algorithm parameter. Then, $(M-1)$ beads,
namely, $next^1 (\alpha)$, $next^2 (\alpha)$, $\,\ldots\,$,
$next^{(M-1)} (\alpha)$ are removed, so that a worm appears with
${\cal I}=\alpha$ and ${\cal M}=\sigma\equiv next^M (\alpha)$. The
acceptance probability for this update is
\beq P_{\rm op} = {\rm min} \biggl \{ 1,\: \frac{C\, \bar{M}\,
N_{\rm bd}\; {\rm e}^{\Delta U  -\mu M\varepsilon }} {
\rho_\circ({\bf r}_{\alpha},{\bf r}_{\sigma},M\varepsilon)}\:
 \biggr \}\; ,
\label{open}
 \eeq
where $\Delta U = U(X)-U(X^\star)$ is the difference between the
$U$-function values for the initial ($X$) and proposed  ($X^{\star}$)
configurations, $N_{\rm bd}$ is the total number of beads in the
initial diagonal configuration equal to the number of particles,
$N$, times the number of slices, $N_{\rm bd}=NP$.  If $M=1$, then
no beads are removed; only the link between the beads $\alpha$ and
$\sigma$ disappears with the appearance of ${\cal I}$ and ${\cal
M}$.
\\
\indent {\it (1b) Close}. This update is only possible if the
configuration is off-diagonal. Let the integer $M\geq 0$ be the
discrete algebraic distance from ${\cal I}$ to ${\cal M}$,
understood as a number of time-slice steps---in the {\it positive
direction} on the $\beta$-cylinder---one has to make to reach the
slice at which ${\cal M}$ is currently positioned, starting from that of ${\cal
I}$. If $M > \bar{M}$ or $M=0$, then the move is
rejected,\cite{note3} otherwise, one proposes to generate a piece
of world line connecting ${\cal I}$ to ${\cal M}$, thereby
rendering the configuration diagonal. If $M>1$, the corresponding
spatial positions of new $(M-1)$ beads, ${\bf r}_{1},{\bf
r}_{2}\ldots ,{\bf r}_{M-1}$, are sampled from the product of $M$
free-particle propagators $\prod_{\nu=1}^{M} \rho_\circ({\bf
r}_{\nu -1},{\bf r}_{\nu},\varepsilon)$, where ${\bf r}_0\equiv
{\bf r}_{\cal I}$ and ${\bf r}_M\equiv {\bf r}_{\cal M}$. The
probability of accepting the move is
\beq P_{\rm cl} = \, {\rm min} \biggl \{ 1, \: \frac{
\rho_\circ({\bf r}_{\cal I},{\bf r}_{\cal M},M\varepsilon)\, {\rm
e}^{\Delta U +\mu M\varepsilon }}{C\, \bar{M}\, N_{\rm bd} }
\biggr \} \; , \eeq
where $N_{bd}$ is the number of beads in the final diagonal
configuration.
In our implementation,  proposed {\it Open} and {\it Close} updates are automatically
rejected whenever the quantity
\[\frac{({\bf r}_{{\cal I}}-{\bf r}_{{\cal M}})^2}{4M\lambda\epsilon} \]
is larger than some (arbitrary) number of order unity, so as to avoid small
acceptance ratios in the close update when the worm ends are far away
in space (in our simulations we set this number equal to 4).
\\
\indent {\it (2a) Insert}. The other way to create an off-diagonal
configuration from a diagonal one, besides {\it Open}, is to seed a new, $M$-link long,
open world line.
 The number of links $1\leq M\leq
\bar{M}$ and the position of ${\cal M}$ in space-time are selected
at random. The spatial positions of the other $M$ beads are
generated from the product of $M$ free-particle propagators. The
move is accepted with probability
\beq P_{\rm in} = \, {\rm min} \{ 1,\; C\, V\, P \, \bar{M}\, {\rm
e}^{\Delta U +\mu M\varepsilon } \}\; , \eeq
where $V$ is the volume of the system, as mentioned above.
\\
\indent {\it (2b) Remove.} The removal of the worm, i.e., of the world line
connecting ${\cal M}$ to ${\cal I}$, is proposed,
provided its algebraic length is $1 \leq M \leq \bar{M}$. (If $M
> \bar{M}$, the proposal is rejected.\cite{note3}) The acceptance
probability for the move is
\beq
 P_{\rm rm} = \, {\rm min} \{ 1,\; {\rm e}^{\Delta U-\mu M
\varepsilon}/CVP\,\bar{M} \}\; . \eeq
At this point, we are in position to discuss the value of the constant $C$, which up to now
we have left undetermined.
A natural choice is
\begin{equation}
C=C_0/VP\,\bar{M} \; , \;\;\;\;C_0 \sim O(1)\; , \label{C0}
\end{equation}
so that the probabilities to {\it Open}, {\it Close}, {\it
Insert}, or {\it Remove} a worm, do not contain macroscopically
large/small factors, and are of order unity at the optimal choice
of $\bar{M}$. Normally, optimal $\bar{M}$ is such that the time
$\varepsilon \bar{M}$ is of the order of the characteristic
single-particle time, which guarantees that the exponentials in
the acceptance probabilities are of order unity, while the
propagators are of the order of the particle number density. This
implies the following scaling:
\begin{equation}
C\; \propto\; {\varepsilon^2 \over V \beta}\; .
\end{equation}
\indent {\it (3a) Advance}. This move advances ${\cal I}$ by a random
number $M$ of slices forward in time. Its implementation is similar to that
of {\it
Insert}. The acceptance probability is
\beq P_{\rm ad} = \, {\rm min} \{ 1,\; {\rm e}^{\Delta U + \mu
M\varepsilon} \}\eeq
\indent {\it (3b) Recede}. Now ${\cal I}$ is displaced backwards in time
(again, in a $\beta$-periodic sense), by erasing $M$ consecutive links;
the number $1\le M \le \bar{M}$ is selected at random. The
acceptance probability is
\beq
P_{\rm re} = \, {\rm min} \{ 1,\; {\rm
e}^{\Delta U - \mu M\varepsilon} \}\eeq
If $M$ turns out to be equal
to or larger than the number of links in the worm, the update is
rejected.\cite{note3}
\\
\indent {\it (4) Swap.} This update is applicable to off-diagonal
configurations only and is illustrated in Fig.~\ref{fig2}. Let ${\cal
I}$ be positioned on the $j$-th slice in the bin $(\vec{\cal
R}_{\cal I},j)$. Consider the $(j+\bar{M})$-th slice (because of
$\beta$-periodicity, this addition is understood modulo $P$) and
create a temporary list, ${\cal L}_{\cal I}$, of all the beads that are contained,
at the slice $j+\bar M$,  in the bins that spatially
coincide with the bin $(\vec{\cal R}_{\cal I},j)$ or with one of  its nearest
neighbors.\cite{note4} Select one bead, $\alpha$, from the list,
with the probability
\beq T_\alpha\; =\; \rho_\circ({\bf r}_{\cal I},{\bf
r}_{\alpha},\bar{M} \varepsilon)/ \Sigma_{\cal I} \; , \eeq
where
\beq \Sigma_{\cal I}\; =\;  \sum_{\sigma \in {\cal L}_{\cal I}}
\rho_\circ({\bf r}_{\cal I},{\bf r}_{\sigma},\bar{M} \varepsilon
)\; \label{Sigma}
\eeq 
is the normalization factor. If any of the beads
$\alpha$, $prev^1(\alpha)$, $prev^2(\alpha)$, $\ldots$,
$prev^{\bar{M}}(\alpha)$ coincides with ${\cal M}$, the move is rejected.
Next, consider the bead $\zeta = prev^{\bar{M}}(\alpha)$ and
identify its bin, $(\vec{\cal R}_{\zeta },j)$. One must now check
whether  the bead $\alpha$ is contained in the bin spatially coinciding
with $(\vec{\cal R}_{\zeta},j)$, or with one of its nearest neighbors;
if that is not the case, the move is rejected.
A second  list is then created, ${\cal L}_{\zeta }$, of beads
contained in the $(\vec{\cal R}_{\zeta },j+\bar{M})$ bin  and in its nearest
neighboring ones.
At this  point,  a set of new positions between beads $\alpha$ and ${\cal I}$
is generated (in the same way as in the {\it Close} move),
${\bf r}_{1}, \ldots , {\bf r}_{\bar{M}-1}$
for the beads
$prev^1(\alpha)$, $prev^2(\alpha)$, $\ldots$, $prev^{(\bar{M}-1)}(\alpha)$,
respectively. One may now rename ${\cal I}$ into
$\zeta $, and vice versa, and re-link beads in the following manner: former
$prev^{(\bar{M}+1)}(\alpha)$ becomes $prev({\cal I})$, etc. As a
result of this world-line reconnection, a piece of world line
between the original bead $\zeta$ and the bead $\alpha$ is erased,
while a new piece of world line appears that connects the former
bead ${\cal I}$ with the bead $\alpha$. The move is accepted with
probability
\beq P_{\rm sw}\; =\; {\rm min} \{ 1,\; {\rm e}^{\Delta U}
\Sigma_{\cal I} /\Sigma_{\zeta} \}  \; . \eeq
Here $\Sigma_{\zeta}$ is defined similarly to $\Sigma_{\cal I}$ in
Eq.~(\ref{Sigma}) using the ${\cal L}_{\zeta }$-list.

The {\it Swap} move generates all possible many-body permutations
through a chain of local single-particle updates. Since no two
particles need be brought within a distance of the order of the
hard core of a typical interatomic potential potential, this move
enjoys a high acceptance rate, similar to that for the {\it
Advance/Recede} updates (we provide a quantitative  example of
acceptance ratios measured in simulations of the two-dimensional
$^4$He liquid below). It must be emphasized that in our algorithm,
unlike in conventional PIMC, arbitrary permutations of identical
particles, as well as macroscopic exchange cycles need not be
explicitly sampled. For, they {\it appear automatically}, if the
physical conditions warrant them. This is because the statistics
of the relative positions for the worm ends is given exactly by
the Green function $G({\bf r}_1,{\bf r}_2, \tau)$.

In a typical {\it sweep}, a worm is created (either by opening an existing closed
world line, or by inserting a new one), advances and/or recedes in imaginary time and
performs a number of swaps, until it finally
closes or is removed. It is easy to convince oneself that, as a result
of this procedure, an exchange cycle
involving a macroscopic number of particles can appear in just one sweep.

Since the complementary pairs are detail-balanced, the final
results do not depend on the global probability of addressing each
pair, so long as the probabilities of addressing each update
within a complementary pair are {\it equal}, which is assumed in
the acceptance probabilities presented above. Otherwise, the
acceptance probabilities should be modified as follows: if the
probability to address update A is $u_A$ and the probability to
address a complementary update is $u_B$ then $P_A \to P_A
(u_B/u_A)$.

As noted in the Introduction, the description of WA given above
is complete even without expansion of the pairwise potential tail
into diagrams (described below). In the original version of the code we did not use
this efficiency enhancing modification of the conventional PIMC
configuration space, and still were able to perform accurate studies
of several hundred atoms. Since expansion into diagrams works only
for the attractive part of the potential it can not be used for purely
repulsive models.

\section{Physical Estimators}\label{mgf}
The statistical estimators for all physical quantities that are
computed in the $Z$-sector, including those of all energetic and
structural properties, are identical with those utilized in
conventional PIMC. We therefore refer the reader to Ref.
\onlinecite{ceperley95} for a detailed discussion of these
estimators. Instead, we focus our discussion here on the estimator
that is a  characteristic feature of the WA, namely that of the
Matsubara Green function.

\subsection{Green function and density matrix estimators}
Since the Green function is sensitive only to relative distances
in time, without loss of generality we can fix $j_{\cal M}=0$ to
simplify the notation. For an extra simplicity, we will also
assume that at least one of the following two statements is true:
(i) The problem is translationally invariant in the coordinate
space. (ii) The quantity of interest is the Green function
averaged over spatial translations,
\begin{equation}
\bar{G}({\bf r},\tau)\; =\; V^{-1}\int d{\bf r}'\, G({\bf r}+{\bf
r}', {\bf r}', \tau) \; .
\end{equation}
In both cases we can formally fix ${\bf r}_{\cal M}=0$. [The
generalization of the treatment to the case of two independent
spatial coordinates is straightforward.]

 Consider the configurations, or path-integral
diagrams, ${\cal D}_{\xi}({\bf r}_{\cal I}, j_{\cal I})$
contributing to the function $g({\bf r}, \varepsilon j)$:
\begin{equation}
g({\bf r}_{\cal I}, j_{\cal I})\; = \; \sum_{\xi}\,  {\cal
D}_{\xi}({\bf r}_{\cal I}, j_{\cal I}) \; .\label{es0}
\end{equation}
The subscript $\xi$ stands for all variables of the diagram
except for the end point space-time positions. That is $\xi$
contains both positions of the beads (in which case the summation
is understood as integration) and a topological structure of the
world lines on the $\beta$-cylinder.

 Comparing the diagrams with
different $j_{\cal I}$ and ${\bf r}_{\cal I}$ (incidentally, it is
precisely this comparison that stands behind the acceptance
probabilities of  {\it Advance} update), we readily see that given
some diagram ${\cal D}_{\xi_0}({\bf r}_0, j_0)$ we can ``upgrade"
it to a diagram ${\cal D}_{\xi }({\bf r}, j)$ with $j>j_0$ by
attaching to ${\cal I}$ a world line piece of $M=j-j_0$ beads.
Specifically,
\begin{equation}
{\cal D}_{\xi}({\bf r}, j) =  {\cal D}_{\xi_0}({\bf r}_0,
j_0)\, R_{j_0 j}({\bf r}_0, {\bf r})~~~~~(j_0<j) \; , \label{es1}
\end{equation}
\begin{equation}
 R_{j_0 j}({\bf r}_0 , {\bf r}) = {\rm e}^{\Delta U + \mu \varepsilon M}
\prod_{\nu=1}^{M} \rho_\circ({\bf
r}_{\nu -1},{\bf r}_{\nu},\varepsilon) \; , \label{es2}
\end{equation}
where $\xi = \{ \xi_0, {\bf r}_1,   {\bf r}_2, \ldots , {\bf
 r}_{M-1}\}$,  ${\bf r}_M\equiv {\bf r}$, and the meaning of $\Delta U$ is the
same as in {\it Advance} update: $\Delta U = U_{\xi_0}-U_{\xi}$,
where $U_{\xi}$ and $U_{\xi_0}$ correspond to the diagrams
$D_{\xi}$ and $D_{\xi_0}$, respectively. Hence, we have
\begin{eqnarray}
g({\bf r}, \varepsilon j) = \sum_{\xi_0}\int d{\bf r}_0 \dots
d{\bf r}_{M-1}\, {\cal D}_{\xi_0}({\bf r}_0, j_0)\, R_{j_0 j}({\bf
r}_0, {\bf r}) = \nonumber \\
\frac{1}{\bar{M}}\!  \sum_{\xi_0, j_0} \! \int d{\bf r}_0 \dots
d{\bf r}_{M-1}\, {\cal D}_{\xi_0}({\bf r}_0, j_0) \, R_{j_0
j}({\bf r}_0, {\bf r}) \delta^{\bar{(M)}}_{j_0 j} \! ,~~~~~
\label{es3}
\end{eqnarray}
where
\begin{equation}
\delta^{\bar{(M)}}_{j_0 j} \, =\; \left\{
\begin{array}{l}
1\; , ~~~ \mbox{if}~~j_0 \in [j-\bar{M}, j-1]  \; , \\
0\; , ~~~\mbox{otherwise}\; .\end{array} \right. \label{es4}
\end{equation}
Though Eq.~(\ref{es4}) is not the final answer yet, it already
contains an important element. It allows one to sample the time
slice $j$ from the adjacent time slices $j_0 \in [j-\bar{M},
j-1]$. The problem with Eq.~(\ref{es4}) is that we still have a
continuous variable ${\bf r}$ while we would like to know the
Green function only at the discrete set of pre-defined points $\{
{\bf r}_p \}$. To proceed further, we utilize (and to a certain
extent generalize) the idea which has been already used in
diagrammatic Monte Carlo.\cite{polaron2000} Suppose we are
interested in $g({\bf r}_p, \varepsilon j)$, where ${\bf r}_p\in
V_p$ is a pre-selected point for collecting statistics and $V_p$
is a 3D volume containing this point. We formally rewrite
(\ref{es3}) at point ${\bf r}_p$ as (below ${\bf r}_M \equiv {\bf
r}_p$)
\begin{eqnarray}
g({\bf r}_p, \varepsilon j)\, = \sum_{\xi_0, j_0}\int d{\bf r}_0\,
{\cal D}_{\xi_0}({\bf r}_0, j_0) \, \delta^{\bar{(M)}}_{j_0 j} \nonumber \\
\times
\, \int d{\bf r}\,  W_{M}({\bf r}_0, {\bf r}) \, \delta^{(V_p)}_{\bf r}
\nonumber \\
\times \,  \int d{\bf r}_1 \ldots d{\bf r}_{M-1}\;
{\prod_{\nu=1}^{M} \rho_\circ({\bf r}_{\nu -1},{\bf
r}_{\nu},\varepsilon) \over \rho_\circ({\bf r}_0,{\bf
r}_p,\varepsilon M)}\; Q\; ,~~~~~~~~ \label{es5}
\end{eqnarray}
\begin{equation}
Q\; =\; (\bar{M} V_p)^{-1}\; {\rho_\circ({\bf r}_0,{\bf
r}_p,\varepsilon M)\over W_{M}({\bf r}_0, {\bf r}) } \; {\rm
e}^{\Delta U +\mu \varepsilon M}\,   \; ,\label{es6}
\end{equation}
\begin{equation}
\delta^{(V_p)}_{\bf r} \, =\; \left\{
\begin{array}{l}
1\; , ~~~ \mbox{if}~~{\bf r} \in V_p  \; , \\
0\; , ~~~\mbox{otherwise}\; .\end{array} \right. \label{es7}
\end{equation}
The value of  $\Delta U$ corresponds to the extra piece of world line
$({\bf r}_0, {\bf r}_1, \ldots , {\bf r}_{M-1}, {\bf r}_p)$. In
principle, $W_{M}({\bf r}_0, {\bf r})$ is an arbitrary
function, but we want it to be positive-definite and normalized
(the integration is over the whole system volume, not just $V_p$),
i.e. to have the meaning of the probability density
\begin{equation}
\int d{\bf r}\, W_{M}({\bf r}_0, {\bf r})\; = \; 1 \; .
\label{es8}
\end{equation}
Moreover, for our purposes the most reasonable choice is simply
\begin{equation}
 W_{M}({\bf r}_0, {\bf r})\; =\; \rho_\circ({\bf r}_0,{\bf
r},\varepsilon M)   \; . \label{es9}
\end{equation}
Now we just need to interpret the relation (\ref{es5}) in terms of
a stochastic process. First, we integrate/sum this relation over
the position of ${\cal M}$ in space/time and compensate for that
by dividing (\ref{es5}) by $PV$. To interpret the first line in
(\ref{es5}), we recall that in accordance with (\ref{ZW}) the
value of ${\cal D}_{{\xi}_0}$ is equal to the probability density
to sample the corresponding diagram of the $G$-sector times the
factor $Z_W/C$. Hence, we can interpret the first line as
averaging---over the ensemble of {\it all} Monte Carlo
diagrams---of the {\it stochastic variable} given by the rest of
the expression times projector $\delta^{(G)}$ times projector
$\delta^{\bar{(M)}}_{(j_{\cal I}-j_{\cal M}), j}$ times $Z_W/CVP$.
The second line says that the evaluation of the stochastic
variable starts with sampling a vector ${\bf r}$ distributed in
accordance with Eq.~(\ref{es9}). The projector
$\delta^{(V_p)}_{\bf r}$ means that if ${\bf r} \notin V_p$, then
the stochastic variable is automatically zero. If ${\bf r} \in
V_p$, then the evaluation procedure continues in accordance with
the third line of Eq.~(\ref{es5}), which we interpret as sampling
$(M-1)$ auxiliary variables, ${\bf r}_1, \ldots , {\bf r}_{M-1}$,
where $M=j-j_{\cal I}+j_{\cal M}$. (In the special case of $M=1$,
auxiliary variables are not sampled.) The auxiliary variables are
sampled from the distribution
\begin{equation}
{\rho_\circ({\bf r}_{0},{\bf r}_{1},\varepsilon)\rho_\circ({\bf
r}_{1},{\bf r}_{2},\varepsilon)\ldots \rho_\circ({\bf
r}_{M-1},{\bf r}_{p},\varepsilon) \over \rho_\circ({\bf
r}_{0},{\bf r}_p,\varepsilon M)}
  \; . \label{es10}
\end{equation}
When these are fixed, the stochastic variable in question---up to
the global pre-factors discussed above---is nothing other than $Q$,
defined by Eq.~(\ref{es6}).

\begin{figure}[t]
\vspace*{-0cm}
\includegraphics[bb=50 250 515 670, width=\columnwidth]{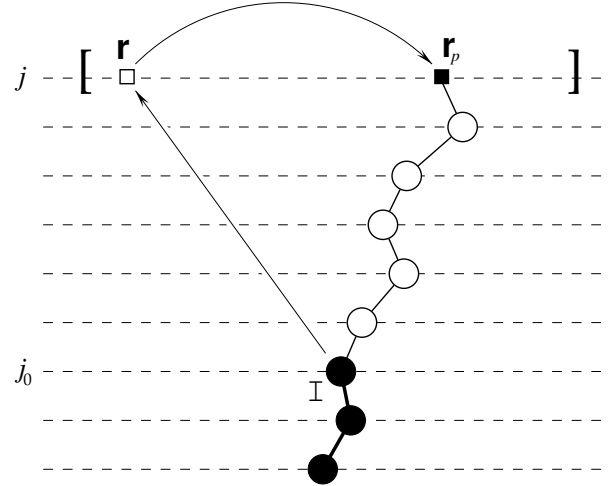}
 \caption{Construction of the
 estimator for the Green function $G({\bf r}_p, \varepsilon j)$,
where without loss of generality
 we assume that ${\bf r}_{\cal M}=0$,  $j_{\cal M}=0$. } \label{fig3}
\end{figure}

Summarizing, we have derived the estimator
\begin{equation}
\langle \, \delta^{(G)} \;  \delta^{\bar{(M)}}_{(j_{\cal
I}-j_{\cal M}), j}\; \delta^{(V_p)}_{\bf r}\, Q \, \rangle_{\rm
MC} \; =\; {CVP\over Z_W}\, g({\bf r}_p, \varepsilon j) \; ,
\label{es11}
\end{equation}
where the variable $Q$ is calculated in accordance with
Eq.~(\ref{es6}) in terms of the auxiliary variables of the
above-described procedure, which we illustrate in Fig.~\ref{fig3},
\[
Q\; =\; \frac{1}{\bar{M} V_p}\; {\rho_\circ({\bf r}_0,{\bf
r}_p,\varepsilon M)  \over \rho_\circ({\bf r}_0,{\bf
r},\varepsilon M) } \; {\rm e}^{\Delta U +\mu \varepsilon M}\; .
\]
The free parameters of the procedure, $\bar{M}$ and $V_p$, are
optimized to yield the best possible convergence.

Our final note is that if one is interested in the
Green function at a certain momentum ${\bf p}$ then
the corresponding estimator does {\it not} require
any elaboration presented above due to extra integration
over ${\bf r}$. Indeed, according to Eq.~(\ref{grj})
we have
\begin{equation}
\langle \, \delta^{(G)} \, \delta_{j,\, (j_{\cal I}-j_{\cal M})}\,
e^{i{\bf p} ({\bf r}_{\cal I}-{\bf r}_{\cal M})} \, \rangle_{\rm MC}
\, = \, {CVP\over Z_W}\, g({\bf p}, \varepsilon j) \; . \label{gpj}
\end{equation}

\section{Enhanced Version: Diagrammatic expansion of the attractive
potential tail}\label{dia}
The WA described in \ref{simplest} can be (and, has been) used for efficient simulations of
systems comprising a few hundred particles. 
This section describes a general procedure that significantly improves
performance of PIMC, through an efficient sampling of the contribution of
the attractive tail of the pair
potential. The trick {\it per se} is not directly related
to the WA, and can be implemented within other PIMC schemes.
Obviously, the modification of the configurational
space it brings about should be adequately taken into account in
the updates. Moreover, in order for the trick to be applicable an attractive tail 
of the interparticle potential must exist, i.e., 
what described below does not apply to purely repulsive interactions, such as
hard-sphere.

Straightforward schemes for calculating the potential energy
exponent of a multi-particle path, $U(X)$, have to deal with a
compromise between accuracy and performance. By a
``straightforward scheme", we mean one
that treats the pair interaction between {\it any} two particles on the
same footing,  irrespectively
of the value of the potential, as long as it is  non-zero, which is the case in all realistic models.
The computational complexity of such schemes scales
linearly with $N$ per single-particle update. To improve on efficiency, one
typically truncates the potential at some distance, thus introducing
a systematic error.

\begin{figure}[t]
\vspace*{-0cm}
\includegraphics[bb=100 250 530 600, width=\columnwidth]{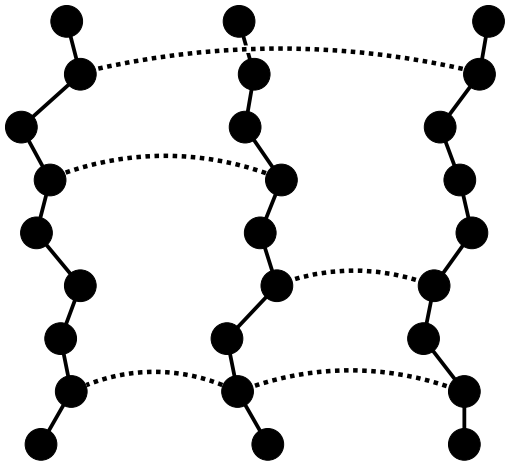}
 \caption{Configuration space with diagrammatic
bonds between the equal-time beads. } \label{fig4}
\end{figure}

There is, however, a way to reduce radically the computational
effort required for accurate treatment of the potential tail,
{\it without} introducing any additional systematic error.  The
only price to pay is a more complex configuration space (see
Fig.~\ref{fig4}) and, correspondingly, additional updates to sample it.

We confine ourselves to the simplest case, i.e., the one in which $U(X)$ is a sum of
pairwise contributions:
\begin{equation}
U\; =\; \sum_{j=0}^{P-1} \sum_{\langle a_jb_j\rangle } \, u({\bf r}_{a_j}-{\bf r}_{b_j})
 \; ,
\label{U_U}
\end{equation}
where beads on the $j$-th time slice are labeled using
subscripts $a_j$ and $b_j$, the notation $\langle a_jb_j \rangle$ stands for
all pairs of beads on a given slice, and ${\bf r}$'s are
the spatial coordinates of the beads. In the simplest\cite{ceperley95} choice for $U$
$u(r)= v(r) \varepsilon$.
Correspondingly,
\begin{equation}
{\rm e}^{-U}\; =\; \prod_j \prod_{\langle a_jb_j\rangle} \, {\rm e}^{-u({\bf
r}_{a_j}-{\bf r}_{b_j})}
 \; .
\label{exp_U}
\end{equation}
An important observation now is that if $|{\bf r}_{a_j}-{\bf
r}_{b_j}| > r_c$, where $r_c$ is some distance, greater than the
radius of the repulsive core of the pair potential, then $u({\bf
r}_{a_j}-{\bf r}_{b_j}) < 0$, and the corresponding pairwise
exponential can be identically split into a sum of two
positive definite terms:
\begin{equation}
{\rm e}^{-u({\bf r}_{a_j}-{\bf r}_{b_j})}\; =\; 1 + \left[\,{\rm
e}^{-u({\bf r}_{a_j}-{\bf r}_{b_j})}-1 \, \right]
 \; .
\label{split}
\end{equation}
Graphically, this decomposition can be represented as follows: Any
two beads within one and the same slice with $|{\bf r}_{a_j}-{\bf
r}_{b_j}| > r_c$,  now may or not share an extra graphical
element, a {\it bond}, see Fig.~\ref{fig4}. The new configuration
space is reminiscent of the Feynman's diagrammatic expansion in
powers of the interaction potential. The absence of a bond between
two beads $a_j$ and $b_j$ represents the first term in
Eq.~(\ref{split}), i.e. unity, and means that the beads at a
distance $|{\bf r}_{a_j}-{\bf r}_{b_j}|> r_c$ do {\it not}
interact. A bond between beads $a_j$ and $b_j$ represents the
second term of Eq.~(\ref{split}). Its relative contribution to the
statistics as compared to the case of no bond has an extra factor
of
\begin{equation}
{\rm e}^{-u({\bf r}_{a_j}-{\bf r}_{b_j})}-1
~~~~~~~~~~~(\mbox{bond factor})
 \; .
\label{bond}
\end{equation}
If one of the beads in the bond is a worm, then the bond factor
is
\begin{equation}
{\rm e}^{-u({\bf r}_{a_j}-{\bf r}_{b_j})/2}-1
~~~~~~~~~~~(\mbox{worm bond factor})
 \; .
\label{bondw}
\end{equation}
Formally, we attribute potential energy exponents to beads, but in reality
they represent interactions between the world line trajectories; since the
worm bead has a trajectory attached to it from one side only, its
potential energy exponent is reduced by a factor of two. Correspondingly,
the two worm beads do not interact and thus ${\cal I}$ can not be connected
by the bond to ${\cal M}$.

The standard MC prescription for sampling the new configuration space
would be to have updates which create and remove bonds.
It is easy to see how a  ``radical gain in performance" can be achieved, as,
statistically, the probability for two beads $a_j$ and
$b_j$ with $|{\bf r}_{a_j}-{\bf r}_{b_j}| > r_c$ to share a bond is
much smaller than unity. Indeed, this probability is proportional
to the bond factor (\ref{bond}), which can be estimated as $\sim
v(r)\varepsilon \ll 1$.
In order to sample the new configuration space, we introduce a pair
of complementary updates that create and remove bonds. This pair
of updates is reminiscent of diagrammatic Monte Carlo updates
that create and delete beads. There is, however, an
important mathematical difference. When creating new beads
one has to seed new continuous variables associated with them.
Bonds are created between existing beads, and formally the new updates
are of the standard Metropolis type.
\\
\indent  {\it Create Bond}. The update is only possible if the
configuration is off-diagonal.  An integer number $M$ is selected
at random within the interval $[0,\bar{M}]$; $\bar{M} < P$. This
number is used to select the first bead, $a_j\equiv prev^{M}({\cal
I})$, in the pair to be connected by the bond. If $prev^{M}({\cal
I})$ is not defined because the world line terminates at ${\cal
M}$, the update is rejected. The second bead, $b_j$, is selected
in two steps. First, within the slice $j$ of the bead $a_j$ we
select a spatial bin, ${\cal B}$. This is done by randomly
generating the bin label from some probability distribution,
$P_{\cal AB}$, which, in general, depends on the distance between
${\cal B}$ and the bin $A$ that contains the first bead $a_j$. (We
discuss a reasonable choice of $P_{\cal AB}$ below.) Let $n_{\cal
B}$ be the number of beads in the bin ${\cal B}$. If $n_{\cal B}
=0$, the update is rejected. If $n_{\cal B} > 0$, we select at
random a bead from the bin ${\cal B}$ and call it $b_j$. If it
happens that $a_j={\cal I}$ and $b_j={\cal M}$ the update is
rejected because physically the world line ends do not interact.
Also, if $a_j$ and $b_j$ are already connected by a bond, or the
distance between the selected beads is smaller than $r_c$, the
update is rejected. Otherwise, we propose to create a bond between
the selected pair of beads, and accept the proposal with the
probability
\begin{equation}
P_{\rm crb}\; =\; {(\bar{M}+1) n_{\cal B} \over  (l_{\rm
bnd}+1)P_{\cal AB} }\,\left[\,{\rm e}^{-fu({\bf r}_{a_j}-{\bf
r}_{b_j})}-1 \, \right]
 \; .
\label{cr_bond}
\end{equation}
Here $l_{\rm bnd}$ is the {\it total} number of bonds in the
initial configuration associated with the beads
${\cal I}$, $prev({\cal I})$, $\ldots$, $prev^{\bar{M}}({\cal I})$
or ${\cal I}$, $prev({\cal I})$, $\ldots$, ${\cal M}$ if
$prev^{\bar{M}}({\cal I})$ is not defined.
The balancing factor $(l_{\rm bnd}+1)$  naturally emerges from the
{\it Remove Bond} update which is complementary to the  {\it
Create Bond}. An additional factor $f=1/2$ in the exponent is necessary only
if one of the beads in the pair is the world line end; otherwise, $f=1$.
\\
\indent  {\it Remove Bond}. The update is only possible if the
configuration is off-diagonal. We list all $l_{\rm bnd}$ bonds associated
with the beads ${\cal I}$, $prev({\cal I})$, $\ldots$, $prev^{\bar{M}}({\cal I})$
or ${\cal I}$, $prev({\cal I})$, $\ldots$,  ${\cal M}$
if $prev^{\bar{M}}({\cal I})$ is not defined.
If $l_{\rm bnd}=0$, the update is rejected. Otherwise, we randomly select
a bond from the list and propose to remove it. The acceptance probability
for the update is
\begin{equation}
P_{\rm rmb}\; =\; { l_{\rm bnd} P_{\cal AB}  \over (\bar{M}+1) n_{\cal
B} }\,\left[\,{\rm e}^{-fu({\bf r}_{a_j}-{\bf r}_{b_j})}-1 \,
\right]^{-1}
 \; .
\label{rm_bond}
\end{equation}
From Eqs.~(\ref{cr_bond})-(\ref{rm_bond}) we realize that an optimal
choice for $P_{\cal AB}$ is based on the interaction potential
between the bin centers
\begin{equation}
P_{\cal AB}\; \propto \;  {\rm e}^{-u(\vec{\cal R}_{\cal B}-
\vec{\cal R}_{A})}-1
\, \approx \, -u(\vec{\cal R}_{\cal B}- \vec{\cal R}_{A})
  \; .
\label{P_B}
\end{equation}

An estimator for the bond contribution to the potential energy,
$U_{\rm bonds}$, is obtained using a standard trick of replacing
$v(r) \to \lambda v(r) $ and then utilizing the identity
\begin{equation}
\langle \, U \, \rangle \; =\;  - {1\over \beta Z}\, { d Z \over
d\lambda }\,  \big|_{\lambda=1} \ ,
\end{equation}
in accordance to which each $Z$-configuration is differentiated
with respect to $\lambda$, and then $\lambda$ is set to unity. The
contribution to the derivative  from the bond factors (\ref{bond})
yields
\begin{equation}
U_{\rm bonds}\; =\;  \frac{1}{\beta}\;\frac{ \langle \,
\delta^{(Z)} \sum_{b} u_{b}/(1-e^{u_b}) \, \rangle_{\rm
MC}}{\langle \, \delta^{(Z)} \rangle_{\rm MC}}\; ,
\label{bond-est}
\end{equation}
where the sum is over all bonds in a current configuration.

\subsection{Worm updates within the diagrammatic bonds}
How does the presence of bonds affect the worm updates? The answer
depends on the updating scenario. The easiest way is to work with
the same updating procedures we had without bonds. The only extra
price is then in having simple constraints on the updates
applicability to a given configuration. Namely, we require that
all the beads being either deleted, or created, or shifted as a
result of worm updates be free of bonds. Special care of bond
factors has to be taken when interconverting regular beads to
worms [see Eqs.~(\ref{bond}) and (\ref{bondw})].

The above requirement is not, in practice, as restrictive as one might think, since the probability of
having no bonds at $\bar{M}$ consecutive beads is of order unity, with the proper choice of parameters discussed in the next
Section.

A brief qualitative discussion of the parameter $r_c$ is in order
here. Formally, $r_c$ can be as small as the size of the repulsive
core of the potential, to guarantee the positive definiteness of
the second term in Eq.~(\ref{split}). It turns out, however, that
this choice is not optimal, because in this case the total
number of bonds per $\bar{M}$ consecutive beads of a world line may grow
large (assuming that $\bar{M}$ is optimized in terms of
the worm updates), and the above-mentioned condition of absence of bonds,
in order for worm updates to be performed, may be satisfied very infrequently;
consequently, the scheme may become inefficient. In the next Section we show that
for helium, $r_c$ is slightly larger than the radius of the first
coordination shell. For best performance,
both $\bar{M}$ and $r_c$ should be simultaneously optimized.

Finally, we note that it is possible to generalize the
diagrammatic procedure to cases when the $e^{-U(X)}$
exponential does not factor into pairwise terms. A great
simplification here comes from the observation that for any
particular choice of $U(X)$, factorization does take place at
least to the first approximation. The correcting terms are then
of the form of close-to-unity three-bead, four-bead (and so
forth) factors. The larger the number of beads in the correcting
factor, the closer the correcting factor is to unity in terms of the
powers of $\varepsilon$. This observation immediately suggests a
recurrent (perturbative) scheme, which consists of ascribing correction factors
to the diagrams, depending on the number of beads in
the bond-connected cluster. The description of such a scheme
goes beyond the scope of the present paper, since we find that
higher-order corrections can be safely neglected for the realistic
choice of $r_c$ discussed in the next Section.

\section{Additional notes on the optimal algorithm parameters
for $^{4}$He.}\label{add}

The separation radius for the diagrammatic expansion is determined
from two conditions which ensure high performance of the algorithm.
For efficient {\it Swap} updates one has to keep the length (in
imaginary time) of the updated trajectory long enough to avoid
proposing large displacements over short time periods.
If $a_0$ is the interatomic distance, then the parameter
$\bar{M}$ must  satisfy the condition
\begin{equation}
\bar{M} \sim \frac{(a_0/2)^2 m }{2\varepsilon } \;, \label{barM}
\end{equation}
which for $^4$He with $a_0=3.5$~\AA ~gives  $\bar{M} \sim 25$
assuming relatively small $\varepsilon = 5\times 10^{-3}
\:K^{-1}$.

Since {\it Swap} and other updates are performed
on trajectory pieces having no bonds on the
corresponding time intervals, it is important to have also
\begin{equation}
\langle \, l_{\rm bnd}(\bar{M})\,  \rangle \, \sim\,  1 \;,
\label{nd}
\end{equation}
where $\langle l_{\rm bnd}(\bar{M}) \rangle$ is the average number of
bonds on the trajectory interval of length
$\bar{M} \varepsilon$. If this number is large, the simple
updating strategy presented above will become inefficient due to
small probability of fluctuations to the state with
$l_{\rm bnd}(\bar{M})=0$. In practice, the computational cost of
calculating acceptance ratios in the strongly-correlated
system is much larger than a simple check for the presence of
bonds; thus, the above condition can be easily
extended to $\langle l_{\rm bnd}(\bar{M}) \rangle \approx 2$.

The number of bonds per time interval can be
readily estimated from the pairwise interaction potential
by using a mean-field estimate for the chemical potential shift
\begin{equation}
\langle l_{\rm bnd}  \rangle\,  \approx\,  \bar{M}\varepsilon n
\int_{r_c}^{\infty} d^d{r}\ v(r)\ g({r}) \, \approx\,
\bar{M}\varepsilon n \int_{r_c}^{\infty} d^dr\ v({r}) \;,
\label{nd2}
\end{equation}
where $g({r})$ is the pair correlation function, which tends to 1
in the ${\bf r} \to \infty $ limit. The last
approximation in Eq.~(\ref{nd2}) is quite accurate for $r$ greater than a
few times $a_0$.
For the Aziz pair potential, $\bar{M}\varepsilon =0.125\:K^{-1}$, and
particle number density $n=0.025\:$\AA$^{-3}$ we have the condition
$\langle l_{\rm bnd} \rangle
\approx 2$ satisfied for
\begin{equation}
r_c \approx 4.2\:\mbox{\AA} \;, \label{Rc}
\end{equation}
in three dimensions. This radius falls in between the first and
second peaks of the pair correlation function, and, roughly
speaking, includes the first coordination sphere of $^4$He atoms.
Thus, on average, not more than 12 neighboring particles are
contributing to the potential energy term in the exponent for the
configuration weight of the trajectory. It should be stressed that
the above set of control parameters is, of course, merely an
approximate guideline. To give a concrete example, for simulations
of liquid helium near the $\lambda$-point at the saturated vapor
pressure (SVP), we find that $r_c=4 \:$\AA ~and $\bar{M}
\varepsilon \approx 0.125$ is a reasonable choice.

The advantage of using sophisticated forms of $U(X)$, lies in the
fact that one may achieve the same accuracy in evaluating the
configuration weight, with smaller number of time slices (see Ref.
\onlinecite{ceperley95} for an exhaustive discussion of this
aspect). Below we discuss the scheme suggested in Ref.~
\onlinecite{chin}, which takes into account the first derivatives
of the interparticle potential (this is the scheme that we have
adopted in all calculations presented below):
\begin{equation}
U(X)=\sum_{j=2k}\frac{2\varepsilon}{3}\: v({\bf R_j}) +
\sum_{j=2k+1}\bigg[ \frac{4\varepsilon }{3}\: v({\bf
R_j})+\frac{\varepsilon^3}{9m}\:F({\bf R_j}) \bigg]  \;,
\label{UX}
\end{equation}
where
\begin{equation}
F({\bf R_j})=\sum_{i=1}^{N}{\bf f}_i^2 \equiv \sum_{i=1}^{N}\bigg(
\sum_{j\ne i}^{N} {\bf f}_{ij} \bigg)^2 \equiv  \sum_{i=1}^{N}\bigg(
\sum_{j\ne i}^{N} \frac{\partial v}{\partial {\bf r}_{ij}} \bigg)^2
\;, \label{FX}
\end{equation}
sums squares of forces acting on particles. The ``force" term is
very important in the repulsive region of the potential and close
to the point where $v(r)$ changes sign. In this region $v(r)$
derivatives are large and their contribution to the configuration
weight may become comparable to the leading linear in
$\varepsilon$ terms on some slices.

Let us separate in Eq.~(\ref{FX}) contributions coming from forces
between close ($r_{ij}<r_c$) and distant ($r_{ij'}>r_c$) pairs (we
label them with indexes 1 and 2 respectively), $F=F_1+F_2+F_{12}$:
\begin{eqnarray}
 F_1&=& \sum_{i=1}^{N}\bigg( \sum_{j\ne
i} {\bf f}_{ij}^{(1)} \bigg)^2 \; ; \;\;\; F_2= \sum_{i=1}^{N}\bigg(
\sum_{j'\ne i} {\bf f}_{ij'}^{(2)} \bigg)^2 \; ; \nonumber \\
F_{12}&=& 2\sum_{i=1}^{N} \bigg( \sum_{j\ne i} {\bf
f}_{ij}^{(1)}\bigg) \cdot  \bigg( \sum_{j'\ne i} {\bf
f}_{ij'}^{(2)}\bigg) \;.  \label{F12}
\end{eqnarray}
If only $F_{1}$ and $F_{12}$ terms were present, the diagrammatic
expansion would be easy to modify to include these terms into the
consideration. The $F_{1}$ term is accounted for directly in the
weight exponent by keeping records of short-range forces acting on
particles and updating them accordingly. We assume that this
procedure is always implemented and discuss below only how to deal
with the force term in the diagrammatic expansion of the potential
tail.

The $F_{12}$ term modifies the diagram weight and acceptance
ratios for the diagrammatic updates. Now, the relative weight
of configurations with and without a bond between distant
(odd) beads $a$ and $b$ is given
by the same formula (\ref{bond}) with
\begin{eqnarray}
u_{ab}=\frac{4\varepsilon}{3}v(r_{ab})+\frac{2\varepsilon^3}{9m}({\bf
f}_a^{(1)}-{\bf f}_b^{(1)})\cdot {\bf f}_{ab}^{(2)} \;.
\label{DFab}
\end{eqnarray}
Since the procedure of keeping track of forces acting on particles
is standard for high-accuracy PIMC schemes, the required
modifications of the scheme are minimal. [Note that short-range
forces are present now in the expression for the bond factor
(\ref{DFab}), and thus their possible effects on bonds
should be accounted for whenever these forces change].

Dealing with the $F_{2}$ term is more cumbersome. Its exact
treatment requires a solution of recursive relations for every
chain of beads connected by the diagrammatic expansion. At this
point we notice that forces between distant particles are {\it
orders of magnitude} smaller than forces acting at short distances
$r<2.6 \:$\AA ~and thus can be safely neglected. Indeed, for
$r_c=4.5\:$\AA ~and $\varepsilon = 5\times 10^{-3} \:K^{-1}$, the
$F_{2}$ term can be estimated to be of order of $10^{-6}$ and thus
has no measurable effect on the simulation results. After all, the
interatomic potential is not even known with this accuracy. We
conclude then that the diagrammatic expansion can be
straightforwardly implemented for the high-accuracy scheme by
omitting the $F_{2}$ term in the configuration weight.

Formally, due to extremely rare statistical fluctuations which
bring two particles at very close distance and result in large
forces ${\bf f}^{(1)}$ acting on them, the sign of $u_{ab}$ in
Eq.~(\ref{DFab}) may change from positive to negative. This, in
turn, will change the sign of the configuration weight if the
corresponding bond is accepted. One can hardly
classify the possibility of such rare events as a ``sign problem"
because the average configuration sign will remain close to unity,
and will not impair the algorithm efficiency. In practice, for
$r_c=4-5 \:$\AA ~and $\varepsilon <0.01\: K^{-1}$ the configuration
sign simply {\it never} changes during the entire simulation.

Finally, one may wonder if the inclusion of the $F_{12}$ term
really helps to achieve better accuracy with smaller number of
slices. If the desired accuracy does not exceed three significant
digits and $r_c$ is kept larger than $4$ \AA ~then the answer is
``No". At this level of accuracy, one may implement the
diagrammatic expansion for the potential tail by ignoring the
force term in the bond factor altogether, i.e. exactly as
described in the previous section. Though we have implemented
schemes with and without the $F_{12}$ term, we did not detect any
difference in final answers when using algorithm parameters
specified above. The $F_1$ term was still kept in the exponent for
more accurate evaluation of the short-range part. Alternatively,
one may choose to deal with the short-range part using the
pair-product approximation of Ceperley and Pollock,
\cite{ceperley95} making use of tables for the two-particle
density matrix calculated for an interaction potential which is
identically zero at distances greater than $r_c$.

\section{Simulation results}\label{resu}

In this Section, we present WA simulation results for bulk liquid $^4$He
in two and three dimensions, with the aim of
demonstrating that superfluid properties and off-diagonal
correlations can be calculated with the WA for very large system
sizes, orders of magnitude larger than accessible to the
conventional PIMC technology. We use the standard interatomic (Aziz)
potential for helium, in an early form for consistency with other calculations.\cite{aziz79}
The reason for our choice of illustrative system, is simply
that the simulation of \he4 in its condensed phase is a
{\it de facto} test bench for new computational many-body techniques.

As mentioned above, we have utilized the form for $U(X)$ suggested in Ref. \onlinecite {chin} for all the calculations for which results are presented here. In both two and three dimensions, we have observed convergence of the kinetic energy estimates (computed with the usual thermodynamic estimator\cite{ceperley95}) using a time step $\varepsilon$=1/640 K$^{-1}$; for all other quantities, four times a value of $\varepsilon$ can be used, and the estimates are seeing to coincide, within their statistical uncertainties, with those extrapolated to the $\varepsilon\to 0$ limit.

\subsection{$^4$He in two dimensions}

\begin{figure}[t]
\centerline{\includegraphics[angle=-90, width=3.6in]{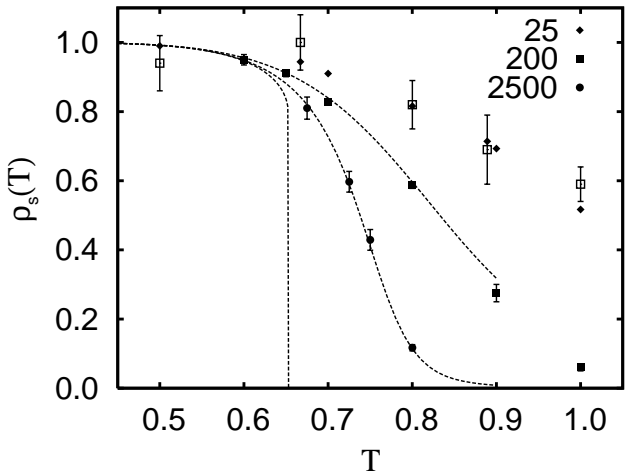}}
\caption{Superfluid fraction $\rho_s(T)$ computed for
2D \he4 on systems with different numbers $N$ of \he4 atoms. The system
density is $n=0.0432$ \AA$^{-2}$. Dashed lines represent fits to the
 numerical data (in the critical region) obtained using the procedure
 illustrated in Ref.~\cite{ceperley89}. The leftmost dashed line is the
 extrapolation to the infinite system. Open squares show results obtained
 in Ref.~\cite{ceperley89} for the same system, with $N$=25. }
\label{fig5}
\end{figure}

We start with the two dimensional case and extend the study
of the superfluid-normal liquid (SF-N) transition first carried out in
Ref.~\onlinecite{ceperley89}, at a density $n=0.0432$ \AA$^{-2}$; we consider system sizes
up to hundred times larger than in the original study.\cite{notebb}
In  Fig.~\ref{fig5} we present data for the superfluid fraction $\rho_s(T)$,
for systems comprising $N=25$, $200$, and $2500$ atoms. Since the
SF-N transition in 2D is in the Kosterlitz-Thouless universality
class,\cite{kt78} with strong (logarithmic) finite-size corrections,
a reliable extrapolation to the thermodynamic limit requires
that simulations be performed for significantly different number of
particles. In the vicinity of the transition point, one may
then employ the asymptotic (in the limit of large distances)
vortex-pair renormalization group (RG) theory to fit the data.
In our study, we used the same RG procedure as in
Ref.~\cite{ceperley89}, which is based on the notion of the vortex
core diameter, $d$ (as a short-range cut-off for RG equations),
and vortex energy, $E_c$, to control vortex density at distance $d$.
Only data in the narrow vicinity of the transition point
0.65 $K$ $\le T \le$ 0.8 $K$ were used in the fitting procedure.
Our estimates for the values of the fitting parameters are $d=8.8\pm
0.5$ \AA\ for the vortex core diameter, and $E$=2.18 $\pm$ 0.04 K
for the vortex energy, which lead to an estimate for the critical
temperature $T_{\rm c}$=0.653$\pm$0.010 K. The vortex diameter
(roughly twice the interatomic distance) turns out to be
comparable to the linear system size $L$ for $N=25$, clearly
showing that  the use of the asymptotic RG analysis is
questionable for such a small system size.
This fact also explains why our result for $T_{\rm c}$ is
significantly different from the previous estimate, 0.72$\pm$0.02 K,
deduced from the $N=25$ data.\cite{ceperley89}

\begin{figure}[t]
\centerline{\includegraphics[angle=-90, width=3.9in]{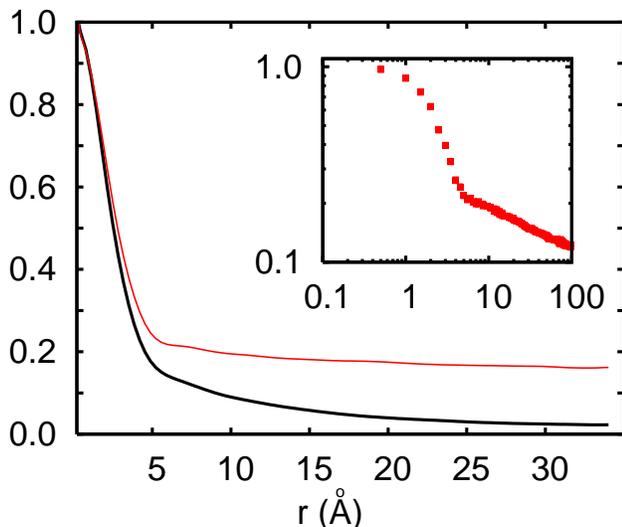}}
\caption{(Color online). One-particle density matrix computed for 2D \he4
at a density $n=0.0432$ \AA$^{-2}$ for a system of 200 atoms, at $T$=0.675 K
(upper curve) and $T$=1.0 K (lower curve). Statistical errors on the curves
are very small, and not shown for clarity.
In the inset,  we present data
 (on a log-log scale) for the N=2500 system at $T$=0.675 K, with clear
 signatures of the Kosterlitz-Thouless behavior in the vicinity of the critical point.
[Reproduced from Ref.~\onlinecite{worm1}.]
  }
\label{fig6}
\end{figure}

In Fig. \ref{fig6}, we show results for the single-particle
density matrix $n(r)$. For 2D helium, this quantity  is expected
to decay to zero at all finite temperatures. In the normal phase, and
far from the critical point, the decay is exponential. At the critical
point, the decay is described by a slow power-law
$ n(r) \sim 1/r^{1/4}$. The same law should be observed in the vicinity
of the critical point up to exponentially large distances.
At low $T$, the exponent in the power law approaches zero.
The data in Fig.~\ref{fig6} are in line with these expectations;
quantitatively, the slope of the $n(r)$ curve on the log-log plot shown
in the inset is close to one quarter.

We wish  to conclude this subsection by giving an example of
typical (without any extensive optimization) algorithm parameters
used for the two-dimensional helium. For the $T=1~K$, $\mu=-1~K$,
and $N=25$ system with $P=200$, $\bar{M}=40$, $C_0=7.5$, and
$r_c=4.05$~\AA ~ the measured acceptance probabilities were
$P_{\rm op}=0.47$, $P_{\rm cl}=0.43$, $P_{\rm in}=0.13$, $P_{\rm
rm}=0.44$, $P_{\rm ad}=0.43$, $P_{\rm re}=0.59$, $P_{\rm
sw}=0.33$, $P_{\rm crb}=0.58$ $P_{\rm rmb}=0.42$.

\subsection{$^4$He in three dimensions}

As mentioned in the Introduction, simulations of bulk 3D liquid helium were among the first
remarkably successful applications of the PIMC method.
\cite{ceperley86} However, previous predictions made
for the superfluid properties were never at the same level
of accuracy as for energetic or structural properties, for reasons
mentioned in the Introduction. In this subsection, we show how the
WA, based on updates described above, eliminates the shortcomings of the existing
PIMC method, by allowing simulations of several thousand atoms
with sufficient precision to determine, for example, the critical temperature
of the SF-N transition at the saturated vapor pressure (SVP)
with accuracy of three significant digits.

\begin{figure}[tbp]
\centerline{\includegraphics[angle=-90, scale=0.32] {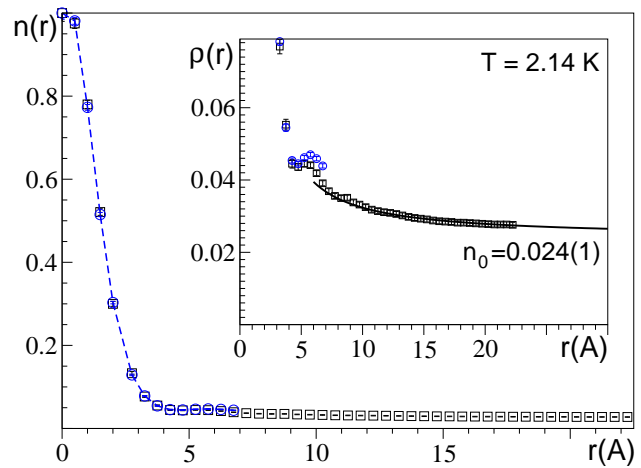}}
\vspace*{-0.4cm} \caption{(Color online). One-particle density
matrix $n(r)$ close to the SVP critical point at $T=2.14~K$
for two system sizes $N=64$ (filled squares) and $N=2048$
(open squares). The solid line is the theoretical prediction
based on long-wavelength phase fluctuations, Eq.~(\ref{ph_fl}).}
\label{fig7}
\end{figure}

In Fig.~\ref{fig7}, we show our data for the density matrix $n(r)$
in the vicinity of the critical point,  at the saturated vapor pressure (at a
temperature $T$=2.14 K and at a  density $n$=0.02198 \AA$^{-3}$)
for system sizes $N=64$ and $N=2048$.
Though the data for small and large system sizes appear nearly
identical in the main plot, the crucial difference
is clearly seen in the inset. The $N=64$ curve terminates
right after the first coordination shell oscillation; the best estimate that can be obtained
of the condensate fraction $n_\circ$ (namely, the asymptotic value to which
$n(r)$ should plateau at long distances) from
this set of data alone, would be about $0.045$. Obviously,
the same coordination shell oscillation prohibits {\it a fortiori} any
reliable finite-size scaling for smaller system sizes, e.g.
using series $N=16,~32,~64$.

In contrast, the $N=2048$ system
is large enough to see the effect of long-wavelength hydrodynamic
phase fluctuations. The Bogoliubov expression for the asymptotic
behavior of $n (r)$ at large distances in the superfluid, is
given by
\begin{equation}
n (r) = n_\circ\ {\rm exp}\biggl [\frac{T}{8\pi\lambda n \rho_s  r  }\biggr ] \;.
\label{ph_fl}
\end{equation}
Since $\rho_s$ is calculated independently, the shape of the
density matrix decay is fixed. The condensate fraction controls
only the overall normalization of the theoretical curve, and this
allows precise extrapolation of the data to the thermodynamic
limit. An example of such extrapolation is shown in
Fig.~\ref{fig7} by the solid line, which predicts $n_0=0.024(1)$
for the condensate fraction $n_\circ$ at $T=2.14~K$---nearly a
factor of two smaller than the $0.045$ estimate obtained on a
64-atom system.

\begin{figure}[tbp]
\centerline{\includegraphics[angle=-90, scale=0.32] {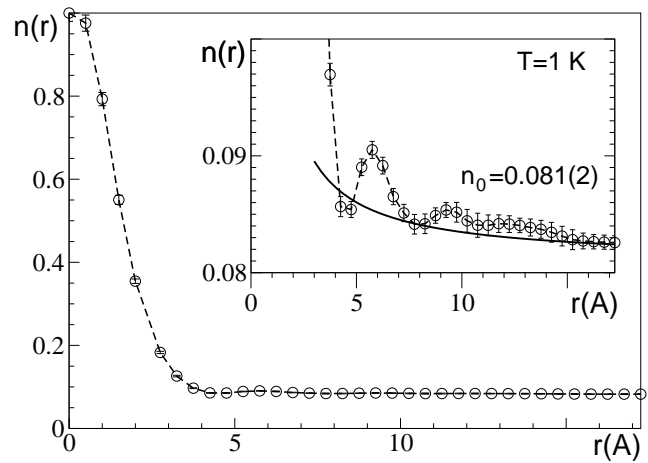}}
\vspace*{-0.4cm} \caption{One-particle density
matrix $n(r)$ for $N=1024$ particles at $T=1~K$ and
SVP pressure. The solid line is the theoretical prediction
based on the long-wave phase fluctuations, Eq.~(\ref{ph_fl}).}
\label{fig8}
\end{figure}

The hydrodynamic correction to the tail of $n(r)$ is less
important at low temperature and for large values of $\rho_s$. For
comparison, in Fig.~\ref{fig8} we present data for a system of
$N=1024$ atoms, at a temperature $T=1~K$ (in this case, the
density is $n$=0.02184 \AA$^{-3}$). In this temperature range,
smaller system sizes can be used to obtain reliable thermodynamic
estimates of $n_\circ$. Our estimate for $n_\circ$ at $T$=1 K is
0.082$\pm$0.002. This is consistent with the existing PIMC
estimate (0.07$\pm$0.01 at $T$=1.2 K, from Ref.
\onlinecite{ceperley95}), obtained on a system of 64 \he4 atoms,
but somewhat above the most recent $T$=0 estimate
(0.069$\pm$0.005, Ref. \onlinecite{moroni04}). Although the
difference is only slightly greater than the combined statistical
uncertainties, it should be noted that finite temperature
calculations are {\it unbiased}, whereas ground state calculations
are based on an input trial wave function. While it is in
principle possible to remove the variational bias associated to
the trial wave function, this may be difficult a goal to achieve
in practice.

Having access to system sizes which allow asymptotic hydrodynamic description
is a necessary condition for determining critical parameters using
finite-size scaling techniques (see, e.g., Ref \onlinecite{Binder}). The idea is
to consider quantities which are determined by system properties
at the largest scales, becoming scale-invariant at the critical point.
For small deviations from criticality, $\delta \to 0$, the dependence
on system size for such quantities is given by
\begin{equation}
R(L,\delta )=f_{R}(L/\xi(\delta) ) \equiv  g_{R} (\delta L^{1/\nu} ) \;,
\end{equation}
where $f_R(x)$ and $g_{R}(x)$ are the corresponding universal
scaling functions ($g_{R}(x)$ is analytic at $x=0$) and $\xi (\delta )$
is the correlation length which diverges at the critical point as
$\xi \propto \delta ^{-\nu }$. For the U(1) universality class in 3D,
the best numerical estimate currently available for the correlation length exponent is
$\nu = 0.6717$.\cite{Camp01}
The intersection of $R(L,\delta )$ curves for different system sizes
provides very accurate and unbiased estimates of critical parameters.

Previous attempts to determine $T_c$ from the scale-invariance of
$R(L)=\rho_sL$  (Josephson relation) have failed.\cite{pollockrunge}
Though the authors of Ref.~\onlinecite{pollockrunge} correctly argue that ``The
statistical uncertainty in the data is too large to accurately
determine $T_c$ from the crossing of these [scaling] two curves",
it seems that coordination shell oscillations also contribute
to several intersections in the 1.6 K $< T <$ 2.4 K interval.

\begin{figure}[tbp]
\centerline{\includegraphics[angle=-90, scale=0.32] {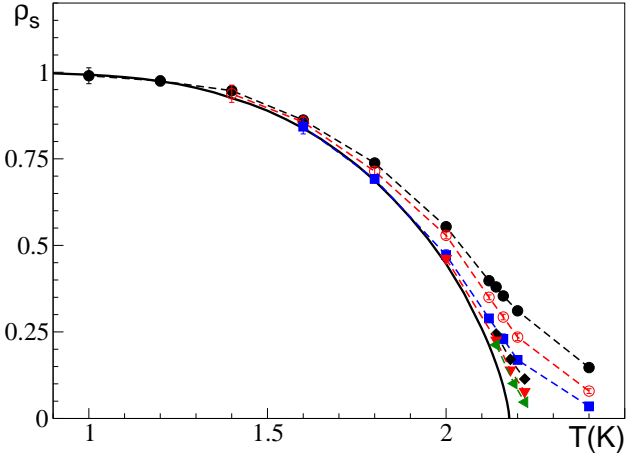}}
\vspace*{-0.4cm} \caption{(Color online). Superfluid fraction
$\rho_s(T)$ as a function of temperature at SVP, computed for
different system sizes, namely $N=64$ (filled circles), $N=128$
(open circles), $N=256$ (filled squares), $N=512$ (diamonds),
$N=1024$ (triangles down), and $N=2048$ (triangles left). The
solid line is the experimental curve.} \label{fig9}
\end{figure}
\begin{figure}[tbp]
\centerline{\includegraphics[angle=-90, scale=0.32] {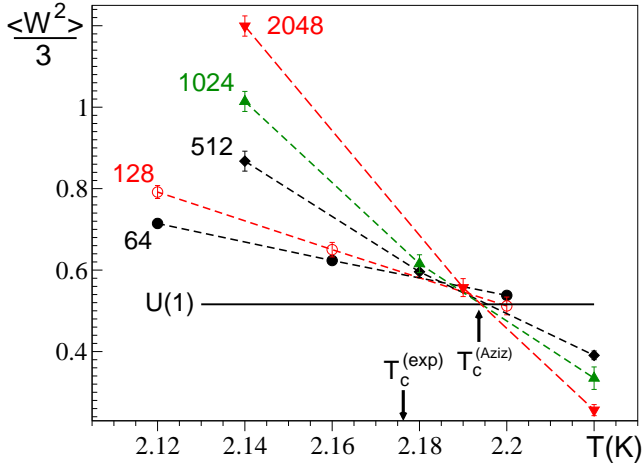}}
\vspace*{-0.4cm} \caption{(Color online). Finite-size scaling plot
for $2\lambda n\rho_sL/T=\langle W^2 \rangle/3$ at SVP. The solid line is the
U(1) universality class value of $0.516(1)$.}
\label{fig10}
\end{figure}
 In Fig.~\ref{fig9}, we show the temperature dependence of the superfluid
fraction for various system sizes. 
After extrapolation to the thermodynamic limit, there is nearly perfect agreement between the numerical
and experimental results.
In order to determine the transition temperature $T_{\rm c}$, we perform
finite-size scaling analysis  of
\[
R(L,T) = \frac{2\lambda n\rho_s L}{T}= \frac{\langle W^2 \rangle }{3},\]
where {\bf W}=$(W_x,W_y,W_z)$
is the winding number.\cite{pollock87} The raw
data are shown in Fig~\ref{fig10}. As an independent check,
we also draw a horizontal line at the known U(1)-universality class
value for winding number fluctuations at the critical point.
\cite{NASA,Camp01} From the intersection of scaling curves (which is seen to
take place at the universal value, within the statistical uncertainties),
we find $T_{\rm c} = 2.193(6)$.  The difference between
this prediction and the experimental value,
2.177 K,  is very small, i.e., simulations of the superfluid
density dependence on system size in the vicinity of the critical
point, do allow us to calculate $T_{\rm c}$ with a relative accuracy of
better than 0.5\%.
There is no reason to expect the Aziz pair potential to reproduce $T_{\rm c}$ 
with much better accuracy, as it was not optimized for this purpose.
It is only a very good approximation to the true interatomic potential
for $^4$He which, in reality, includes irreducible forces acting
between three and more particles.\cite{moroni2000}   

\begin{figure}[tbp]
\centerline{\includegraphics[angle=-90, scale=0.32] {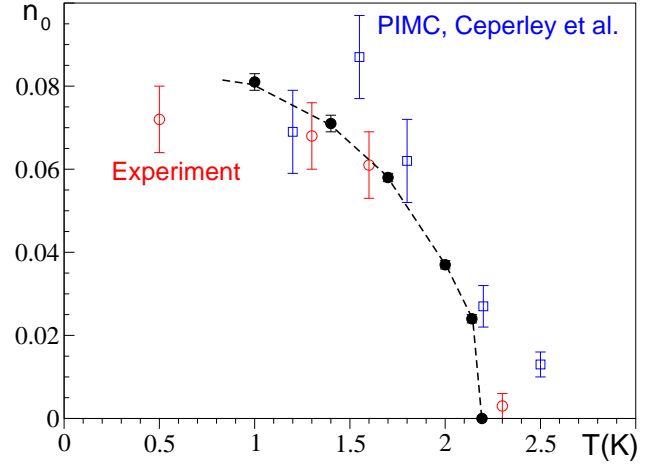}}
\vspace*{-0.4cm} \caption{(Color online). Condensate fraction
at the saturated vapor pressure. The dashed line is used to guide
the eye.}
\label{fig11}
\end{figure}

Condensate density has been consistently one the most difficult properties to
compute and measure for helium. Shown in Fig.~\ref{fig11} are
available experimental data \cite{Glyde} along with the previous
PIMC results \cite{ceperley95} and new estimates obtained in the
present study, extrapolated to the thermodynamic limit as
explained above. With new technology we substantially reduce
theoretical uncertainties on predictions of $n_o$ at the SVP.

We conclude by mentioning that we have also obtained estimates for all other
standard thermodynamic quantities, such as the kinetic energy, pair correlation function etc.
Our data are generally consistent with those of existing calculations; specifically, our $T$=1 K results
are indistinguishable, within statistical uncertainties, from
those yielded by numerically exact ground
state methods (see, e.g., Ref. \onlinecite{pnr}).

\begin{figure}[tbp]
\centerline{\includegraphics[angle=-90, scale=0.30] {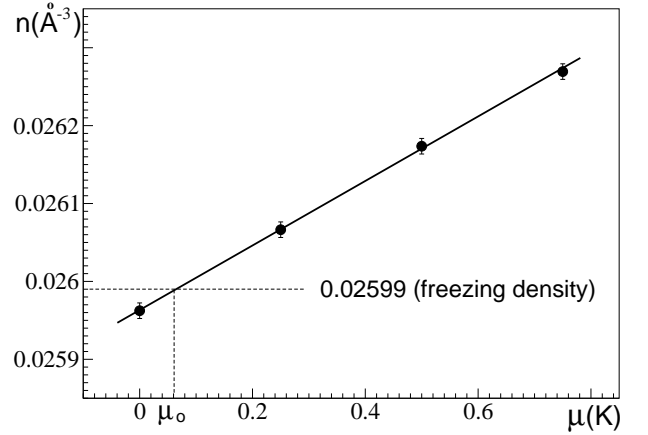}}
\vspace*{-0.4cm} \caption{(Color online). Liquid density as a function of
chemical potential at low temperature $T=0.25$ K. The slope of the
solid line is deduced from the simulated values of compressibility
averaged over four points shown in the plot. The critical value of the
chemical potential deduced from this plot is estimated as
$\mu_\circ =0.06 \pm 0.04$. }
\label{fig12}
\end{figure}

\section{Discussion and Conclusions}\label{conclusions}
We have introduced a new Worm algorithm, affording an accurate
PIMC study of strongly correlated Bose systems. Illustrative
results of numerical simulations of the superfluid transition in
liquid \he4 have been presented, for system sizes two orders of
magnitude larger than what is accessible to conventional PIMC. It
should be stressed that such an advance cannot be simply
attributed to the availability of faster computing facilities than
back in the days when the first PIMC simulations of  liquid \he4
were carried out. Rather, the WA decisively overcomes the most
important limitation of conventional PIMC, namely the exponential
inefficiency with which long permutation cycles are sampled, in
the thermodynamic limit (a limitation acknowledged by
practitioners of  PIMC\cite{bernu04}).

We have also described a procedure, based on ideas of Diagrammatic Monte Carlo,
which allows one to enhance significantly the scalability of the computational
scheme, without compromising on the accuracy of the calculation.

The new methodology has already been applied to the study of the supersolid
phase of helium,\cite{superglass} for which access to large system sizes is
crucial; it can also be expected to have broad impact on a wide variety of
strongly correlated quantum many-body systems. It should be mentioned, that
the efficiency with which long permutation cycles can be sampled using the WA,
also significantly impacts the convergence of calculations of the superfluid
properties of {\it finite} systems, such as quantum clusters.\cite{mezzacapo}

There are other advantages to this new method, chiefly the fact
that it is fully grand canonical, and that allows for the
calculation of the Matsubara Green function, a quantity that can
not e computed with any other existing QMC technique (in
continuous space). An immediate application of these last two
aspects, is the calculation of chemical potentials and excitation
gaps. Consider, for instance, the calculation of the chemical
potential at the liquid-solid transition line at low temperature,
$\mu_\circ$. The study of $G(k=0, \tau)$ in the solid phase can be
found in Ref.~\onlinecite{vacancies}.

\begin{figure}[tbh]
\centerline{\includegraphics[angle=-90, scale=0.32] {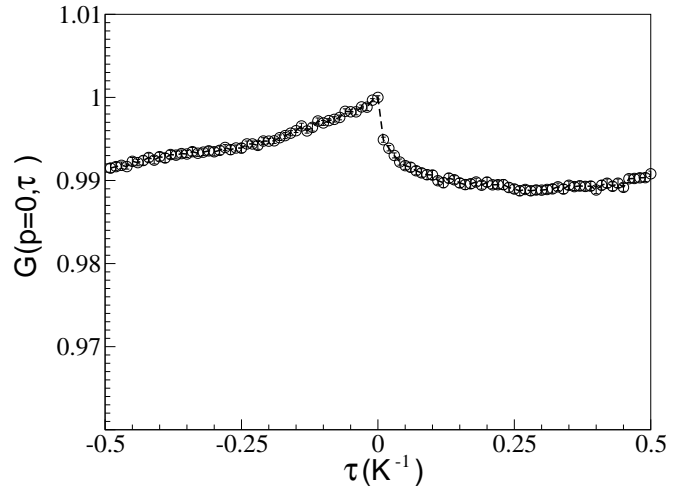}}
\vspace*{-0.4cm} \caption{Zero-momentum Green function for the
superfluid state of $N\approx 1000$ atoms at $T=1$ K, $\mu =
-7.35$ K, and density $n=0.02184$ \Am3~. It is normalized to unity
at the origin. Note the small vertical scale: The
$\tau$-dependence is a finite-size effect.} \label{fig13}
\end{figure}

The value of $\mu_\circ$ can be determined from the $n(\mu)$ curve
and known freezing density for the liquid which is at $n=0.02599$ \Am3.
In Fig.~\ref{fig12} we show data points for four values of the chemical
potential and the position of the freezing density. All simulations
were performed at $T=0.25$ K, and with the particle number
$N \approx 800$. Apart from average density we also calculated
statistics  of particle number fluctuations to obtain the compressibility
of the system from
$\kappa = dn/d\mu = \langle (N-\langle N \rangle)^2 \rangle/TV$. It provides
an independent check for consistency and convergence of the data.
In Fig.~\ref{fig12} the slope of the solid line is obtained
from the average value of $\kappa$ for all data points shown.
From this set of data we deduce that the critical value of the chemical
potential  as
\begin{equation}
\mu_o = 0.06 \pm 0.04 \;.
\label{mu0}
\end{equation}
Though one can deduce this parameter from the ground state
energy as a function of particle number using $\mu = dE/dN \approx E(N+1)-E(N)$,
such simulations can not be performed for large system sizes
with accuracy significantly better then few K.

The zero-momentum Matsubara Green function $G(k=0, \tau)$ for the
superfluid state at $T=1$ is shown in Fig.~\ref{fig13}. In the
macroscopic limit,  $G(k=0, \tau)$ is $\tau$-independent, being
equal to the total number of condensate particles. We show this
plot to support our previous claim that in the superfluid state
the two worm ends perform a random walk on large distances, and
come close within the reach of the {\it Close} and {\it Remove}
updates once per sweep.

\section{Acknowledgments}
This work was supported by the National Aero and Space
Administration grant NAG3-2870, the National Science Foundation
under Grants Nos. PHY-0426881, NSF PHY-0456261, by the Sloan
Foundation, and by the Natural Science
and Engineering Research Council of Canada under grant G121210893. NP gratefully 
acknowledges hospitality and support from the Pacific Institute of Theoretical Physics, Vancouver (BC).


\begin{thebibliography}{99}
\bibitem{notebf}
The WA can be applied to Fermi systems as well; however it will be
affected by the {\it sign} problem. Although we do not discuss the
fermion problem here, we note that approximate schemes such as the
``fixed-node approximation" (see, for instance, Ref.
\onlinecite{ceperley96}) can be easily incorporated in the WA.
\bibitem{notev}
Note that this assumption on the potential is made for simplicity only;
the applicability of the WA does not depend on it, i.e., more general interactions can be
studied without altering the computational scheme in any significant way.
\bibitem{note1} Using standard importance sampling, one may
control statistics of particle number fluctuations, and perform
canonical ensemble measurements.
\bibitem{note3} Since balance equation deals only with {\it accepted}
moves, one can propose a move without {\it a priori} checking that
the current configuration is consistent with the proposal.---If it
is not, the move is automatically rejected after the
proposal, with no calculation needed.
\bibitem{note4} The specific definition  of ``nearest
neighbor" is not relevant for the generic description of the move.
\end{thebibliography}
\end{document}